# Shaping the Future of Social Media with Middleware

December 2024

**Editors**: Luke Hogg (Foundation for American Innovation) and Renée DiResta (McCourt School of Public Policy at Georgetown University)

**Contributors**: Francis Fukuyama (Freeman Spogli Institute for International Studies at Stanford University), Richard Reisman (Foundation for American Innovation), Daphne Keller (Cyber Policy Center at Stanford University), Aviv Ovadya (AI & Democracy Foundation), Luke Thorburn (King's College London), Jonathan Stray (Center for Human-Compatible AI at UC Berkeley), and Shubhi Mathur (Clavata.ai)

**Symposium Participants**: Min Li Chan, Tracy Chou, Elena Cryst, Paige Collings, Alex Feerst, Zach Graves, Evan "Rabble" Henshaw-Plath, Danny O'Brien, Chand Rajendra-Nicolucci, Chris Riley, Mitch Stoltz, Rose Wang, and Dave Willner

**Acknowledgements**: Robert Bellafiore, Rhonda Bonneville, Alissa Cooper, Joshua Levine, Mike Masnick, and Richard Whitt

This paper is the product of a symposium jointly held by the Foundation for American Innovation and the Stanford Cyber Policy Center in April 2024. The goal of this symposium and subsequent paper is to build upon and operationalize the concept of middleware as it was proposed by the Stanford Working Group on Platform Scale led by Dr. Francis Fukuyama in 2020. While this paper synthesizes numerous conversations and presentations with symposium participants and draws heavily on the works of others, the content of this paper represents the views of the contributors alone. The contents of this paper should not be attributed to individual symposium participants, who may or may not agree with some or all of the views expressed herein.

# Executive Summary

*Middleware*, third-party software intermediaries between users and platforms, offers a promising solution to counter the concentrated power of social media platforms. The term has referred to a variety of technologies and systems over the years, including third-party provider tools that platforms themselves use internally. In this paper, we focus on "middleware" in the form of open, third-party products and services that are composable—meaning, with multiple providers available to be mixed and matched for specific use cases—and which offer user agency over the selection process and overall experience.

Our analysis centers on middleware's potential to transform two common social media experiences that are often the focus of controversy, user dissatisfaction, and political debates: curation, which involves selecting and organizing information to shape what content is emphasized or deemphasized in a digital environment, and moderation, which addresses harmful content and compliance with platform policy. By providing users with greater control over these experiences, middleware promotes a more user-centric, democratic online sphere. It enables users to choose from competing providers and algorithms, offering a flexible architecture as an alternative to both centrally controlled, opaque platforms and an unmoderated, uncurated internet. Middleware has the potential to provide greater choice around the content individual users see, and to address over-moderation concerns. By decentralizing control and enhancing user autonomy, middleware may also help to reduce the potential for abuse of power by platforms, fostering a more just and equitable digital ecosystem.

The success of open middleware presently hinges on the adoption and cooperation of established major platforms. For middleware to thrive in the present largely centralized environment, platforms must permit third-party services to operate and enable users to choose between them. However, the growing rise of federated platforms, such as Mastodon and Bluesky, and the increasing participation of major platforms in the fediverse (e.g., Meta's Threads), creates new opportunities for the development and adoption of middleware as an integral part of the user experience. These emerging ecosystems prioritize user choice and both horizontal and vertical interoperation, allowing for community-driven moderation tools and enhanced user control of the social media experience.

In light of this shift in technology and adoption, the Foundation for American Innovation and the Stanford Cyber Policy Center convened a gathering of experts in April 2024 to explore the implications of advancing middleware adoption. The group included technologists, entrepreneurs, policymakers, activists, civil society leaders, academics, and independent researchers. Over a day of deliberations, participants examined middleware's potential as a transformative force to reshape the digital public sphere, enhance user agency, and address persistent challenges in content curation and moderation. They also considered the trade-offs of middleware and the negative externalities it might create, as well as the technological, regulatory, and market barriers that could either support or hinder its implementation.

This report synthesizes those discussions and outlines key considerations for the future of middleware:

- **Transformative Potential**: Middleware's role in decentralizing power, enhancing user agency through flexible architectures, and facilitating community-centered, democratic online ecosystems.
- **Risks and Trade-Offs**: The potential for middleware to exacerbate polarization, facilitate harassment or persistent harmful content, entrench echo chambers, fragment norms, or undermine privacy.
- **Technical Feasibility**: The challenges of developing scalable, secure, and interoperable middleware systems that can integrate with diverse platform architectures while protecting user privacy.
- **Economic and Market Dynamics**: The opportunity to avoid replicating misaligned incentives from centralized social media platforms, and instead to create sustainable middleware business models.
- **Public Policy and Regulatory Frameworks**: The importance of governance mechanisms or standards to ensure privacy, accountability, and alignment with societal values while fostering innovation.

We hope that this contribution serves as a thoughtful assessment of middleware's promise and its complexities, and offers a roadmap for policymakers, technologists, and stakeholders to navigate this emerging landscape.

# **Introduction**

*"We shape our tools and thereafter our tools shape us." – John M. Culkin*[1]

The digital age has fundamentally altered the nature of the public square. In prior eras, public squares were communal spaces where information exchange and social norms evolved through collective participation. Today, the metaphor often describes online communities and interactions shaped within a digital ecosystem dominated by a handful of private platforms, each driven by its own incentives. This shift represents a profound change in how we communicate, share information, and reach consensus. A significant portion of our interactions occurs on these large social media platforms, which heavily influence the information we see and our norms of engagement, giving them enormous power over public discourse.

---

[1] John M. Culkin S.J., "A Schoolman's Guide to Marshal McLuhan," *The Saturday Review,* March 1967, p. 70, https://webspace.royalroads.ca/llefevre/wp-content/uploads/sites/258/2017/08/A-Schoolmans-Guide-to-Marshall-McLuhan-1.pdf.

In 2020, the Stanford Working Group on Platform Scale considered the potential harms to society posed by the scale, scope, and power exhibited by concentrated digital platforms.[2] It concluded that the scale and concentrated power of current digital platforms cause not just well-documented economic harms, but social and political harms as well: as information curators and moderators, platforms steer user attention, and influence democratic discourse and deliberation. Moreover, as private entities, major platforms face few checks on their power—there are no elections for leadership, nor referendums on policy. In considering potential remedies for minimizing the harms caused by this concentrated private power, the working group brought widespread attention to a compelling argument for "middleware" as a way to reshape the organizational structures of social media, and rebalance power between platforms and users.[3]

The working group defined middleware as "software and services that would add an editorial layer between the dominant internet platforms and consumers."[4] In this paper we focus on *open, third-party middleware*, or third-party software and services that act as delegated user agents, intermediate between users and platforms, and are "open" in the sense that they are accessible and adaptable, allowing for third-party integration without undue restrictions. For example, middleware can offer independent fact-checking services for tweets, the ability to switch between distinct and tailored news feeds on Facebook, or filters that adjust Amazon search results to prefer domestically produced sustainable products, to name just a few applications. While the emphasis of this report is on middleware agents for social media platforms that enable many-to-many communication, much of the discussion is applicable to applications like e-commerce and search engines that also involve feeds or recommenders—and soon, perhaps, to AI agents operating on behalf of users.

The power of middleware lies in its flexibility. It enables an opening of the organizational structure of social media platforms to discovery, experimentation, and innovation. By introducing a competitive layer of service providers, middleware dilutes platform control over user feeds and attention, and empowers users to exert more direct control over their online experience. In the realms of content curation and moderation, middleware enables trusted consumer- or community-centric organizations to act as user agents, potentially creating more tailored and meaningful interactions. This flexibility is potentially valuable not only at the individual level but also at the societal level, where it might foster a more balanced and democratic digital environment more akin to a true public square. As middleware reduces platforms' ability to selectively amplify or silence voices, it might also be used to address moderation concerns without relying on more restrictive structural or content-based laws and regulations, which may create their own negative externalities.

---

[2] Francis Fukuyama et al., "Report of the Working Group on Platform Scale," Stanford Cyber Policy Center, November 17, 2020, https://cyber.fsi.stanford.edu/publication/report-working-group-platform-scale.
[3] Barak Richman and Francis Fukuyama, "How to Quiet the Megaphones of Facebook, Google and Twitter," *Wall Street Journal*, February 12, 2021, https://www.wsj.com/articles/how-to-quiet-the-megaphones-of-facebook-google-and-twitter-11613068856.
[4] Ibid.

Middleware offers a promising solution for decentralized, user-empowering, and community-driven online interactions.[5] However, to fully realize a thriving middleware marketplace within the major platform ecosystem, there must be a level of platform accommodation and buy-in. Platforms must allow open, third-party middleware services to operate, and must enable users to choose between them. Some of the most powerful centralized platforms have taken moderate steps towards more openness—meaning, a lack of restrictions on the ability of users to integrate middleware. But most have not yet embraced middleware for reasons we will discuss, and at least one (X/Twitter) has backtracked.

Recent technological shifts toward decentralization and federated social media platforms, however, have created new opportunities for experimentation with middleware. With mainstream-owned, emerging platforms such as Meta's Threads built to be compatible with the "fediverse,"[6] and protocol-based federated platforms such as Mastodon, Bluesky, and Nos gaining in popularity, promising new markets for middleware are emerging on social media that are architected from the start to enable more user choice and interoperation.[7] Policy shifts are underway as well: the European Union's Digital Markets Act's portability and interoperability mandates are likely to promote the growth of a middleware ecosystem, and several proposals are percolating in the United States that could further accelerate this trend. We are potentially at a crossroads, where users may choose between continued centralized social media experiences or participation in a digital environment that may have fewer features or presently smaller networks, yet gives more agency to individuals and communities. This paper aims to explain the promises and trade-offs of the latter.

In its original 2020 report, the Stanford Working Group on Platform Scale laid out a positive vision for what middleware might enable, but left several questions to be answered: What should the role and function of middleware be? What sort of business model for middleware providers might induce innovators to build middleware products and services? How should public policy be aligned to support the development of a market for middleware?

In an effort to answer those questions, to review more recent work, and to provide concrete recommendations for users, builders, and policymakers, the Stanford Cyber Policy Center and the Foundation for American Innovation jointly hosted an April 2024 symposium of leading thinkers around decentralization, interoperability, and middleware. This report is the outcome of that conversation, and of ongoing conversations since. Building on the work of the Stanford Working Group and many other academics, it aims to operationalize the vision of a social media ecosystem that encourages more user agency through middleware.

---

[5] See Fukuyama et al., "Report of the Working Group on Platform Scale."
[6] The fediverse—a portmanteau of "federated" and "universe"—is a decentralized network of independently hosted servers or "instances" that interoperate through a common set of protocols, allowing users on different servers to interact with each other seamlessly.
[7] "What Is the Fediverse?" Meta, June 25, 2024, https://about.fb.com/news/2024/06/what-is-the-fediverse/.

We begin by briefly examining the current social media landscape and the growing concerns over platform power. We then explain the concept of middleware and discuss its potential to increase user agency, transform content curation and moderation, and address current shortcomings around reputation and trust on social media. Next, we analyze the market dynamics for middleware, exploring the importance of strategic alignment between platforms and middleware providers; consider possible business models; and identify potential avenues for the evolution of the market. Finally, we turn our attention to public policy, examining how existing legal frameworks might impact the development of middleware, and evaluating various public policy shifts that might improve middleware's chances for success.

# **Background**

Online pioneers envisioned a world wide web imbued with the qualities of openness and collaboration. As activist John Perry Barlow put it in 1996, the rise of the open, standards-based internet could usher in a world "where all may enter without privilege or prejudice accorded by race, economic power, military force, or station of birth … where anyone, anywhere may express his or her beliefs, no matter how singular, without fear of being coerced into silence or conformity."[8] A multitude of community social systems such as The WELL, ECHO, and Bulletin Board Systems (BBSs) had emerged, enabling socialization and human connection. But by the 1990s, centralized, proprietary entrants such as CompuServe, Prodigy, and AOL began to compete with these community systems. These proprietary services and their successors increased adoption of user-generated content spaces by making it easier for the average person to create and participate, and for new collaborations to emerge.[9]

As adoption grew, the internet became both ubiquitous and impactful; anyone could create content, and some of the content was controversial. Policymakers and the general public began to grow concerned about the accessibility of explicit and harmful content, particularly for minors. As the potential harms of wide propagation of unfiltered, any-to-any speech came under scrutiny, the question of how to think about the responsibility of platforms that carried user-generated content became a focus of lawmakers. Congress attempted to impose order with the Telecommunications Act of 1996, which included both speech restrictions subsequently struck down by the Supreme Court in the seminal *Reno v. ACLU* case and a safe harbor for platform moderation commonly referred to as Section 230. This surviving provision established two important rules: one which "confers general immunity to Internet service providers and websites by preventing them from being treated as publishers of content created by third parties," and a second which "guarantees that the immunity is not abrogated by the online intermediary's effort's to screen out objectionable content."[10] In other words, the platforms themselves were not liable for content posted by their users, but they also had the right to curate and moderate it. Without this indemnity, social media as we know it would likely not have become a largely open "public square" for user-created content; lawsuits would have put the providers out of business.[11]

---

[8] John Perry Barlow, "A Declaration of the Independence of Cyberspace," accessed through the Electronic Frontier Foundation, February 8, 1996, https://www.eff.org/cyberspace-independence.
[9] E.g., blogging services made it easier to publish written content without having to know HTML, and YouTube made publishing video content as easy as uploading a file. New social groups formed, based on commonalities that were not limited by geography.

Over the subsequent decades from the 1990s to the 2020s, waves of "social networks" and "social media" companies swelled and faded. From GeoCities and SixDegrees in the 1990s, to Friendster and MySpace in the 2000s, and Facebook, YouTube, and Twitter in the decades since, online platforms competed to become central gathering places where users spent their time. They became spaces for people worldwide to share life updates, find new friends, discover culture, create content, and build communities around topics of interest. As they grew, these centralized private platforms came to largely capture the market for online social interaction, including by buying and assimilating upstarts (as Facebook did with Instagram, and Google did with YouTube), "sherlocking"[12] features (as Instagram did with Reels and Twitter did with Spaces), or shutting potential competitors down through litigation (as Facebook did with Power.com).

It quickly became clear that social platforms were more than just places to connect around hobbies and interests. Large networks of people and amplification tools were also a source of political power. Platforms had become critical tools for organizing, spreading messages, and shaping public opinion. Grassroots political and social-good movements, such as the Arab Spring of 2011, captured the imaginations of activists and alarmed some governments as people parlayed online energy into offline action.[13] Authoritarian governments came to embrace the tools as well, using social media to manipulate their own publics, or to interfere in the politics of rivals: military leaders in Myanmar incited a genocide against an ethnic minority[14] and the Kremlin's online trolls surreptitiously inserted themselves into regional conflicts, then American politics.[15] By the mid-2010s, scholars argued, "platforms themselves amassed sufficient power that they could potentially sway an election, either as a matter of deliberate choice or as a result of being unwittingly manipulated by other political, state or non-state actors."[16] The massive scale of modern platforms granted them unparalleled access to vast audiences, similar to the dominance of television networks in the mid-20th century; this potentially enabled them to mold public opinion through proprietary, opaque content feeds and content policy decisions. Furthermore, the vast amount of information online means that users are practically dependent on platforms to moderate, rank, and recommend content.

---

[10] Jeff Kosseff, "Defending Section 230: The Value of Intermediary Immunity," *Journal of Technology Law & Policy*, vol. 15, issue 2, article 1 (2010), https://scholarship.law.ufl.edu/jtlp/vol15/iss2/1/.

[11] Jeff Kosseff, "Social Media Is a Mess. Government Meddling Would Only Make It Worse," *New York Times*, December 14, 2023, https://www.nytimes.com/2023/12/14/opinion/supreme-court-social-media.html.

[12] "Sherlocking" refers to the practice by a platform or technology company of creating its own version of a popular third-party application or feature, thereby diminishing the original application's relevance or market share. See G.F., "You've Been Sherlocked," *The Economist,* July 13, 2012, https://www.economist.com/babbage/2012/07/13/youve-been-sherlocked.

[13] Zeynep Tufekci, *Twitter and Tear Gas: The Power and Fragility of Networked Protests* (New Haven: Yale University Press, 2017).

[14] Paul Mozur, "A Genocide Incited on Facebook, with Posts from Myanmar's Military," *New York Times,* October 15, 2018, https://www.nytimes.com/2018/10/15/technology/myanmar-facebook-genocide.html.

Today, a handful of social media companies and platforms significantly shape how we access and spread information, connect with others, conduct transactions, and earn a living. The business models of the largest social media platforms rely on collecting vast quantities of information about their users. Sometimes criticized as "surveillance capitalism,"[17] the instrumentation of user behavior and leveraging of massive data analysis to determine user preferences at a minimum turns platforms into effective curators and recommenders of online content. Platforms rank feeds, amplify popular posts or topics, recommend accounts to follow and groups to join, and moderate content.

As they do so, they play a significant role in shaping public discourse, influencing user emotions, and steering user attention. The audience is not entirely passive, of course, but unaccountable private power is ripe for abuse. Social science research has identified social media platforms as a contributor to affective polarization[18] and as a risk to democracy, markets, personal data, and digital rights.[19] In 2016, a "techlash" movement rose to prominence in response to concerns about online harms and the negative externalities of platform content decisions.[20] Activists advocated for weakening the private power of Big Tech, for better accountability frameworks, and for a return to the internet's foundational principles and value-based business models that prioritize mutual benefit over extraction. Many called attention to platform business models, arguing that financial incentives drove platforms toward promoting inflammatory or otherwise attention-grabbing content.[21]

---

[15] U.S. Senate Select Committee on Intelligence, *Russian Active Measures Campaigns and Interference in the 2016 U.S. Election, Volume 1: Russian Efforts Against Election Infrastructure with Additional Views* (2019), https://www.intelligence.senate.gov/sites/default/files/documents/Report_Volume1.pdf.

[16] See Fukuyama et al., "Report of the Working Group on Platform Scale."

[17] Shoshana Zuboff, *The Age of Surveillance Capitalism: The Fight for a Human Future at the New Frontier of Power* (New York: Public Affairs, 2019).

[18] Petter Törnberg, "How digital media drive affective polarization through partisan sorting," *Proceedings of the National Academy of Sciences* (October 10, 2022), https://www.pnas.org/doi/abs/10.1073/pnas.2207159119.

[19] Philipp Lorenz-Spreen *et. al.,* "A Systematic Review of Worldwide Causal and Correlational Evidence on Digital Media and Democracy." *Nature Human Behavior* 7, 74–101 (2023), https://doi.org/10.1038/s41562-022-01460-1.

[20] John S. and James L. Knight Foundation and Gallup, "Techlash? America's Growing Concern With Major Technology Companies*,*" March 11, 2020, https://knightfoundation.org/reports/techlash-americas-growing-concern-with-major-technology-companies/.

[21] Sinan Aral, *The Hype Machine: How Social Media Disrupts Our Elections, Our Economy, and Our Health—and How We Must Adapt* (Crown Currency, 2020).

One way to mitigate the risk of unaccountable private power is regulation. Globally, governments have pushed platforms for more transparency in their curation and moderation processes and attempted to create frameworks for accountability for online harms such as election interference or unchecked disinformation campaigns.[22] However, this process is often *ad hoc*, inconsistent, and reactive, and enforcement remains a challenge. Mandating that platforms filter harmful content, promote accurate information, and respect freedom of expression may sound simple, but it has proven enormously difficult in practice. In the United States, legislative and regulatory efforts are fraught with challenges, particularly because the First Amendment limits the government's ability to dictate how platforms curate or moderate content.[23]

On the self-regulatory front, even platforms that view ethical content curation and moderation as beneficial to their business face challenges in finding the right balance between protecting free expression and minimizing harm. Moderation significantly shapes the user experience on a social media platform; most people do not want to spend time in spaces overrun with hate speech, harassment, and spam, even if such content is legally permissible.[24] Therefore, platforms have responded by attempting to remove or reduce such content. However, different communities have different criteria and tolerance thresholds for what is or is not acceptable. Politicians with regulatory power are disinclined to see messages associated with their party down-ranked or labeled. Many civil society organizations and influential individuals have also raised concerns about the potential abuse of unaccountable private power in the context of moderation, highlighting issues of bias, cultural insensitivity, and censorship.[25] A combination of inconsistent and opaque enforcement of moderation policies, and the politicization of moderation in some locales (such as the United States) has led to a crisis of legitimacy, with platforms facing criticism from all sides.[26]

As "techlash" concerns grew, prominent individuals and media within the tech world began to advocate for more open, modular, and distributed alternatives to centralized social media platforms. Some argued for decentralization and federated platforms; tech writer Mike Masnick outlined a vision of rearchitecting social media to focus on "protocols, not platforms" as a way to maximize free expression and improve the marketplace of ideas.[27] Others argued for improvement within the confines of centralized platforms; one early idea, for example, suggested that Facebook should allow outside providers to curate news feeds.[28] Users might subscribe to the *New York Times* feed, the MrBeast feed, the Joe Rogan feed, the Rachel Maddow feed, or the Republican National Committee feed, via middleware.[29]

---

[22] Some of these laws, such as Singapore's Code of Practice for Online Safety or India's Information Technology (Intermediary Guidelines and Digital Media Ethics Code) Rules, apply within one country. The European Digital Services Act covers the European Union.
[23] See *Moody v. NetChoice, LLC*, No. 22–277 (2024).
[24] Emily A. Vogels, "The State of Online Harassment," Pew Research Center, January 13, 2021, https://www.pewresearch.org/internet/2021/01/13/the-state-of-online-harassment/.
[25] Christina Pan et al., "Algorithms and the Perceived Legitimacy of Content Moderation," Stanford University: Human-Centered Artificial Intelligence, December 2022, https://hai.stanford.edu/sites/default/files/2022-12/HAI%20Policy%20Brief%20-%20Algorithms%20and%20the%20Perceived%20Legitimacy%20of%20Content%20Moderation.pdf.
[26] Jason Koebler and Joseph Cox, "The Impossible Job: Inside Facebook's Struggle to Moderate Two Billion People," *Vice*, August 23, 2018, https://www.vice.com/en/article/xwk9zd/how-facebook-content-moderation-works.

The Stanford Working Group on Platform Scale raised the concept to broader awareness with its 2020 white paper proposing the development of middleware technologies that sit between users and platforms in order to mitigate the problems of unaccountable private power, bias, cultural insensitivity, and censorship. By enabling users to choose from a suite of services such as feed composers, recommenders, and moderation services provided by third parties, middleware can bring more legitimacy to platform governance, offer more choices to users, and reconcile competing priorities. It can be designed to align with local laws and culture, diverse contexts, and ever-evolving norms, enabling sovereign countries and distinct communities alike to tailor content filters and moderation standards. It can also introduce competition and innovation into markets currently dominated by a few major players. However, as we will discuss in the next section, there are also meaningful trade-offs, including risks of weakened moderation of illegal content, that arise when putting control in users' hands.

## **Middleware as a Solution**

Digital platforms such as Google, Facebook (Meta), Instagram, TikTok, Twitter (now X), and Apple play a significant role in the daily lives of billions of people across the globe. While the socioeconomic benefits of these platforms are substantial, their concentrated power also grants them significant—and potentially dangerous—control over the flow of information.[30] Social science research suggests that social media has had negative effects on social cohesion,[31] and may have weakened democratic processes and institutions that rely on informed, responsible, and collaborative citizens.[32] Economically, concentrated platform power also allows companies to stifle competition.[33]

---

[27] Mike Masnick, "Protocols, Not Platforms: A Technological Approach to Free Speech," Knight First Amendment Institute at Columbia University (August 21, 2019), https://knightcolumbia.org/content/protocols-not-platforms-a-technological-approach-to-free-speech. See also Mike Masnick, "Protocols Instead Of Platforms: Rethinking Reddit, Twitter, Moderation And Free Speech, *TechDirt*, July 17, 2015, https://www.techdirt.com/2015/07/17/protocols-instead-platforms-rethinking-reddit-twitter-moderation-free-speech/.
[28] Daphne Keller, "Who Do You Sue? State and Platform Hybrid Power Over Online Speech," Hoover Institution, Aegis Series Paper no. 1902 (2019), https://www.hoover.org/sites/default/files/research/docs/who-do-you-sue-state-and-platform-hybrid-power-over-online-speech_0.pdf.
[29] Michael J. Coren, "Facebook Needs to Hand Over its Algorithm If It Really Wants to End Fake News," *Quartz,* December 6, 2016, https://qz.com/847640/facebook-needs-to-hand-over-its-algorithm-if-it-really-wants-to-end-fake-news.
[30] See Fukuyama et al., "Report of the Working Group on Platform Scale."
[31] Sandra González-Bailón and Yphtach Lelkes, "Do Social Media Undermine Social Cohesion? A Critical Review," *Social Issues and Policy Review*, vol. 17 (December 2022), p. 155–180, https://spssi.onlinelibrary.wiley.com/doi/10.1111/sipr.12091.

The 2020 Stanford Working Group paper laid out the value proposition of middleware:

> Middleware's primary benefit is that it dilutes the enormous control that dominant platforms have in organizing the news and opinion that consumers see. Decisions over whether to institute fact-checking, remove hate speech, filter misinformation, and monitor political interference will not be made by a single CEO but will instead be controlled by a variety of informed and diverse intermediaries. … Additionally, middleware facilitates competition. It offers a new and distinct layer of potential competition for consumer loyalties and opens a pathway for innovations in managing information, including commercial information that might benefit firms otherwise disadvantaged by the platforms' business models.

Middleware has the potential to improve the balance of power between platforms and users, providing individuals with greater control over their online experiences. In this section, we examine its capacity to facilitate more effective content curation and moderation, bolster reputation and trust, revitalize social media innovation, and ensure that interactions in the digital space are more contextual, reliable, and transparent. We also consider the technical feasibility of middleware services, and discuss potential concerns about the social impact of middleware.

## Increasing User Agency in Curation

One of the most significant benefits of middleware is its capacity to empower users by giving them greater control over their online experience. By enabling more granular control over the content they see and interact with, middleware allows users to tailor their social media environment to fit their needs and preferences, redistributing power and enabling a more direct, active role in shaping their online interactions. On centralized platforms in particular, this shift promotes a more democratic digital sphere where users have a say in managing their interactions, and where users are less vulnerable to platforms wielding opaque power for nefarious purposes.

One area in which middleware could significantly increase user agency is via replacements or enhancements for platform-controlled content curation and recommendation algorithms. A firehose of information assembled into a bottomless feed of posts is often overwhelming for users. In response, platforms devised curation algorithms, ranked feeds, and personalized recommenders that are responsible for selecting which content the user should see, from whom, and in what order. However, the incentives of the "attention economy"[34] have resulted in curatorial and recommendation functions that are misaligned with user or societal interest (for example, the promotion of sensational or rage-inducing content keeps people engaged on the platform, which may be good for the platform but has deleterious personal and social impact). Multiple studies have suggested that poorly considered recommendation and feed ranking algorithms may inadvertently promote actively harmful or radicalizing content, or send users down "rabbit holes."[35] However, efforts to study algorithmic feeds struggle with lack of data access and transparency.

---

[32] Joshua Tucker et al., "Social Media, Political Polarization, and Political Disinformation: A Review of the Scientific Literature," Hewlett Foundation (March 2018), https://papers.ssrn.com/sol3/papers.cfm?abstract_id=3144139.
[33] Lina Khan, "The Separation of Platforms and Commerce," *Columbia Law Review,* vol. 119, no. 4 (2019), https://columbialawreview.org/content/the-separation-of-platforms-and-commerce/.

Some platforms do offer users a choice of feed formats, typically either a reverse-chronological feed ("Recent Posts") or an algorithmically ranked feed ("For You"). In a reverse-chronological feed, posts from followed accounts appear in order of recency, with the latest post at the top. Algorithmically ranked feeds are far more variable: sometimes they prioritize recency, sometimes topicality, sometimes accounts with whom a user frequently interacts. Content might also be recommended based on patterns from users with similar behavioral histories. The exact criteria for ranking are often opaque. TikTok, for example, is known for its "black box" algorithm, which surfaces videos the platform predicts will engage the user without requiring a follow.[36] Other platforms, like Facebook and Instagram, have increasingly adopted this approach, with "Unconnected Posts"—content from accounts the user hasn't followed—making up a growing share of feed content.[37]

Some users find the algorithmically curated *status quo* serendipitous; others find it creepy. In order to make "good" personalized suggestions, platforms collect and process significant amounts of data about user behavior in ways that increasingly implicate questions of privacy. Users have limited ability to provide feedback to algorithmic curators. They can indicate that they don't want to see a particular account or piece of content again, but beyond choosing which accounts to follow and rejecting individual content suggestions, they have little proactive control. They must rely on the platform to infer that certain types of content are broadly undesirable to them. A growing chorus of social media users express frustration with algorithmic unpredictability or perceived bias even as they continue to use the platforms.[38] Social media platforms, after all, are increasingly where most socialization happens, and where most news and information is consumed.[39]

---

[34] First coined by Nobel laureate Herbert Simon, the "attention economy" refers to an information rich system in which attention for consuming information is a more limited resource than the information itself. In digital markets, the attention of users has become a limited and valuable resource to be commodified.
[35] Jonathan Stray, Ravi Iyer, and Helena Puig Larrauri, "The Algorithmic Management of Polarization and Violence on Social Media," Knight First Amendment Institute at Columbia University (August 22, 2023), https://knightcolumbia.org/content/the-algorithmic-management-of-polarization-and-violence-on-social-media.
[36] Ben Smith, "How TikTok Reads Your Mind," *New York Times,* December 5, 2021, https://www.nytimes.com/2021/12/05/business/media/tiktok-algorithm.html.
[37] Meta, "Widely Viewed Content Report: What People See on Facebook: Q1 2024 Report," 2024, https://transparency.meta.com/data/widely-viewed-content-report/.
[38] Monica Anderson, "Americans' Views of Technology Companies," Pew Research Center*,* April 29, 2024, https://www.pewresearch.org/internet/2024/04/29/americans-views-of-technology-companies-2/.

A market of middleware providers specializing in curation could empower users to select among feed-ranking algorithms, perhaps subscribing to multiple and toggling between them, allowing users to fine-tune content preferences, avoid unwanted or irrelevant material, or shape their online experiences according to their tastes, interests, and moods.[40] Greater transparency in curation would also help demystify the algorithmic filtering process, increasing user awareness of how algorithmic choices shape their information environment.

By facilitating this "freedom of impression," middleware has the potential to not only enhance user agency but also promote a more personalized and positive user experience.[41] Users or independent middleware providers could potentially create their own feeds, then share them with the platform community so that others can subscribe. Experimental efforts to rank feeds on centralized social media platforms, such as the Gobo project, have offered users the ability to create "lenses" to filter and sort posts and surface particular types of content such as cute animals or professional news sources.[42] Protocol-based social network Bluesky now features custom feeds and "algorithmic choice" (along with composable moderation as described in the next section) that enable users to create and subscribe to custom feeds. Some of the most popular presently include "Art" (featuring artists' posts) and "Blacksky" (a feed aggregating posts by self-declared Black users).[43]

Future middleware-based agent services might enable users to delegate curation to branded providers of their choosing—for example, subscribing to feeds assembled by the *New York Times*, Fox News, *Sports Illustrated*, or their local church or community center. Users would not need to understand the intricacies of the curation process; they would simply choose services they trust to align with their interests, potentially even paying a premium for this assurance.[44] This option to subscribe to differentiated feeds could help reduce the prevalence of sensationalism, clickbait, and ragebait that often arise from engagement-based ranking incentives on centralized platforms, which offer only one or two feed views and create a winner-take-all attention competition. It also reduces the concern that centralized platforms can surreptitiously put their thumb on the scale in favor of political parties, positions, or candidates by preferentially ranking content along ideological lines—an allegation that Meta, TikTok, Alphabet, and Twitter/X have each faced.[45]

---

[39] "News Platform Fact Sheet," Pew Research Center, September 17, 2024, https://www.pewresearch.org/journalism/fact-sheet/news-platform-fact-sheet/.
[40] Nick Couldry, "Resonance, Not Scalability," Ash Center for Democratic Governance and Innovation, Harvard Kennedy School (February 2, 2024), https://ash.harvard.edu/resources/resonance-not-scalability/.
[41] Richard Reisman, "The Internet Beyond Social Media Thought-Robber Barons," *Tech Policy Press,* April 22, 2021, https://www.techpolicy.press/the-internet-beyond-social-media-thought-robber-barons/.
[42] Spencer Lane, "Gobo 2.0: All Your Social Media in One Place", Public Infrastructure blog, Nov 9, 2022, https://publicinfrastructure.org/2022/11/09/gobo-2-0-all-your-social-media-in-one-place/.
[43] Martin Kleppmann et al., "Bluesky and the AT Protocol: Usable Decentralized Social Media*,*" Arxiv, February 5, 2024, https://arxiv.org/pdf/2402.03239.
[44] Richard Reisman, "Making Social Media More Deeply Social with Branded Middleware," Smartly Intertwingled, October 13, 2024, https://ucm.teleshuttle.com/2024/10/making-social-media-more-deeply-social.html; Richard Reisman and Chris Riley, "Delegation, or, The Twenty Nine Words that the Internet Forgot," *Tech Policy Press*, February 28, 2022, https://techpolicy.press/delegation-or-the-twenty-nine-words-that-the-internet-forgot/.

Transforming Content Moderation

Curation, feed ranking, and recommendation systems aim to surface desirable content to an individual user. By contrast, content moderation policies and enforcement mechanisms attempt to demote or remove undesirable or harmful content. Moderation affects not only individual users but also "bystanders" and the broader platform environment. Middleware options for content moderation could empower users with more granular control over their online environments, fostering a more harmonious experience that respects the diverse needs and values of different user communities.

Presently, content moderation on centralized platforms largely involves top-down judgment calls about what materials and behaviors are acceptable. Some lines are straightforward: certain types of content, such as child sexual abuse materials, are illegal. But many content moderation concerns are far less cut-and-dry, and platforms set their own rules for addressing them—often to the chagrin of specific segments of the user base. For instance, some platforms allow pornography and nudity, while others ban it outright. There is a spectrum of tolerance across centralized social media platforms for hate speech, misinformation, bots, spam, gore, sex, nudity, and other types of content that might cause harm, offense, or discomfort. This range of content, sometimes called "lawful-but-awful," is perceived differently across cultures and communities.[46] Because there are few universal norms, it is challenging to establish universal rules.

Nonetheless, most platforms attempt to moderate lawful-but-awful content with a largely one-size-fits-all approach, sometimes adapting policies to regional contexts. Governance rules are set and enforced by the company, in a model sometimes referred to as "customer service" moderation:[47] platforms respond to user reports, attempting to balance individual rights with the health of the overall user community.[48] The specifics may change depending on the values of the leadership team, business considerations, or in response to external groups "working the referees," yet ultimately the platform sets the rules for all of its users.

---

[45] Emily A. Vogels, Andrew Perrin, and Monica Anderson, "Most Americans Think Social Media Sites Censor Political Viewpoints," Pew Research Center, August 19, 2020, https://www.pewresearch.org/internet/2020/08/19/most-americans-think-social-media-sites-censor-political-viewpoints/; Jerrold Nadler, "Letter to Jim Jordan," August 12, 2024, https://democrats-judiciary.house.gov/uploadedfiles/2024-08-12_jn_to_jdj_-_x_grok_misinfo.pdf; and U.S. House Judiciary Committee, *Reining in Big Tech's Censorship of Conservatives,* (2020), https://judiciary.house.gov/sites/evo-subsites/republicans-judiciary.house.gov/files/2020-10/2020-10-06-Reining-in-Big-Techs-Censorship-of-Conservatives.pdf.

[46] Daphne Keller, "Lawful but Awful? Control over Legal Speech by Platforms, Governments, and Internet Users," *University of Chicago Law Review Online,* June 28, 2022, https://lawreviewblog.uchicago.edu/2022/06/28/keller-control-over-speech/.

[47] Ethan Zuckerman and Chand Rajendra-Nicolucci, "From Community Governance to Customer Service and Back Again: Re-Examining Pre-Web Models of Online Governance to Address Platforms' Crisis of Legitimacy," *Social*

Top-down governance models such as this are often criticized for lack of transparency, accountability, and redress. Resulting resentment has contributed to a politicization and crisis of legitimacy in content moderation among some communities.

However, there is another model for content moderation: community moderation, in which governance is handled by members of the community rather than by professional moderators. This approach is often perceived as more legitimate by the community.[49]

In the early days of the social internet, community members helped to set the norms and rules of their online spaces. That model has continued in some spaces today. For example, Reddit's hybrid approach means that the platform itself sets some top-down ground rules, but otherwise a largely decentralized approach empowers subreddits to moderate themselves. Individual moderators in these smaller spaces enforce the community's standards. They can remove content, mute or ban users, and establish custom policies aligned with Reddit's guidelines such as prohibitions on "not-safe-for-work" content, personal information sharing, and upvote solicitation.[50] Mods are also empowered to create additional content policies that define acceptable content and use for their subreddits: no posting pictures of dogs in the cat photo subreddit. Similarly, on federated social media platform Mastodon, local instance rules are set by the server owner.[51] Mastodon is not a single service, but a federated network of community-centered instances based on open-source software that can be modified as each community manager desires. Users can choose to join servers in line with their values and content moderation preferences; instances can "defederate" from other servers that host content objectionable by their standards, and their users will not see it.[52]

---

*Media + Society,* vol. 9, issue 3 (July-September 2023), https://journals.sagepub.com/doi/10.1177/20563051231196864.

[48] Kate Klonick, "The New Governors: The People, Rules, and Processes Governing Online Speech," *Harvard Law Review*, vol. 131, issue 6 (April 2018), https://harvardlawreview.org/print/vol-131/the-new-governors-the-people-rules-and-processes-governing-online-speech/.

[49] Ethan Zuckerman and Chand Rajendra-Nicolucci, "Let the Community Work It Out: A Throwback to Early Internet Days Could Fix Social Media's Crisis of Legitimacy," NiemanLab, October 25, 2023, https://www.niemanlab.org/2023/10/let-the-community-work-it-out-a-throwback-to-early-internet-days-could-fix-social-medias-crisis-of-legitimacy/.

[50] Spandana Singh, "Everything in Moderation: An Analysis of How Internet Platforms Are Using Artificial Intelligence to Moderate User-Generated Content," *New America*, last updated July 22, 2019, https://www.newamerica.org/oti/reports/everything-moderation-analysis-how-internet-platforms-are-using-artificial-intelligence-moderate-user-generated-content/case-study-reddit/.

[51] "Moderation Actions," *Mastodon,* accessed December 3, 2024, https://docs.joinmastodon.org/admin/moderation/.

[52] Alan Z. Rozenshtein, "Moderating the Fediverse: Content Moderation on Distributed Social Media," *Journal of Free Speech Law*, vol. 3, issue 217 (2023), https://www.journaloffreespeechlaw.org/rozenshtein2.pdf.

While not all platforms are structurally designed in the way that Reddit or other persistent group-focused communities are, middleware could help to incorporate community moderation principles into more centralized or newsfeed-focused platforms. Just as users might subscribe to curation middleware that ranks content, they might subscribe to community-designed or third-party tools that filter it based on a variety of signals (likes, shares, comments, etc.) from one or more specific communities.

Bluesky, for example, has begun to experiment with giving users greater agency over content moderation in a Twitter-like environment.[53] Its main instance, bsky.social, has top-level moderation standards that address spam and other significant issues. However, it additionally offers users the ability to share blocklists and labeling frameworks which tag content or users with labels such as "hate speech," "porn," "gore," and other disfavored content types.[54] Users have the option to manage how material with those labels shows up in their feeds, choosing whether to hide, display warnings about, or ignore the content. Users can choose which labelers to follow and report content to those moderators specifically, who in turn determine the appropriate label for their subscriber community. This customization allows users to tailor their experience according to their comfort levels with various types of content, even as the platform takes a more expansive view toward free expression.

It is important to note that early experiments in community labeling on Bluesky have yielded mixed results; community moderation teams have struggled to manage the volume of labeling requests, and there have been some instances of community moderation disagreement leading to the wholesale disbanding of specific labeler efforts.[55] Trust and safety work is complex, and significantly time consuming; as we will discuss in the section on Markets for Middleware Development, in order for user-controlled solutions to scale, middleware moderation offerings will require a viable economic model. However, Bluesky's effort is a noteworthy and ambitious experiment in using middleware to actively incorporate the user community into this process, increasing their agency over their social environment.[56]

It might be argued that centralized platforms could themselves provide more direct user controls for moderating content without the messy complications of an open middleware infrastructure.

---

[53] Jay Graber, "Composable Moderation," Bluesky, April 13, 2023, https://bsky.social/about/blog/4-13-2023-moderation.

[54] "Labels and Moderation," Bluesky, accessed October 8, 2024, https://docs.bsky.app/docs/advanced-guides/moderation.

[55] Sarah Perez, "Bluesky Addresses Trust and Safety Concerns around Abuse, Spam, and More," *TechCrunch,* September 18, 2024, https://techcrunch.com/2024/09/18/bluesky-addresses-trust-and-safety-concerns-around-abuse-spam-and-more/.

[56] Mike Masnick, "Why Bluesky Remains the Most Interesting Experiment In Social Media, By Far," *TechDirt,* March 27, 2024, https://www.techdirt.com/2024/03/27/why-bluesky-remains-the-most-interesting-experiment-in-social-media-by-far/ .

Some platforms have taken steps toward openness in this direction, partly because heavily centralized moderation is so costly and its legitimacy is so often questioned intensely. Individual users and group moderators on some platforms have access to keyword blocking tools. More collaboratively, Twitter/X introduced Community Notes,[57] which allows users to collaboratively append fact-checks and corrections to posts containing misleading information, and previously allowed API access to some third-party content moderation tools such as Block Party, which enabled users to block and mute users or keywords *en masse*.[58] However, platform support for third-party offerings has been intermittent; a 2015 effort to offer shared exportable blocklists, Block Together, lost the ability to operate after Twitter shut down the ability to export the lists.[59] Furthermore, even centrally managed global providers have struggled to keep up with the wide diversity of distinct user and community needs and contexts that billions of global users want addressed, and efforts to explore more granular control are worthwhile.

## Improving Reputation and Trust Online

Concerns about content moderation and curation reflect deeper underlying issues of reputation and trust on social media platforms. Human discourse offline involves not only individual agency (which determines what individuals we speak with and listen to, and what groups we participate in) and information flows, but reputation assessments and social mediation processes. Reputation signals help us decide which individuals and communities to pay attention to, both individually and collectively, and communities and institutions shape what we know and come to believe.[60] The incentives of centralized social media platforms have not always aligned with providing mechanisms to fulfill these needs and support constructive sensemaking.[61] Middleware may be transformative in doing so.

Social media platforms democratized the dissemination of information, turning everyone into content creators and reducing the power of gatekeepers. However, this shift also enabled the proliferation of viral misinformation, disinformation, rumors, and divisive propaganda. While these issues are not new, the mechanisms that communities once relied on to evaluate information in slower, more localized information environments have not been effectively recreated online. The explosion of voices, while a net positive, has made it increasingly difficult to discern accurate information and identify authoritative sources.

---

[57] "Community Notes: A Collaborative Way to Add Helpful Context to Posts and Keep People Better Informed," X, accessed October 8, 2024, https://communitynotes.x.com/guide/en/about/introduction.

[58] After Twitter put access to its API behind a paywall, Block Party pivoted to a privacy focused middleware in 2024 known as Privacy Party. See Tracy Chou, "Coming to Terms with the Messy Spectrum of Online Speech," Block Party, January 10, 2023, https://www.blockpartyapp.com/blog/coming-to-terms-with-the-messy-spectrum-of-online-speech/.

[59] Jacob Hoffman-Andrews, "Update 2021-01-18: Block Together Is Now Fully Shut Down," *Blocktogether.org*, June 16 2020, https://blocktogether.org.

[60] Richard Reisman, "Three Pillars of Human Discourse," Tech Policy Press, October 24, 2024, https://www.techpolicy.press/three-pillars-of-human-discourse-and-how-social-media-middleware-can-support-all-three/, "New Logics for Governing Human Discourse in the Online Era," Centre for International Governance Innovation (April 25, 2024), https://www.cigionline.org/publications/new-logics-for-governing-human-discourse-in-the-online-era/; "A New, Broader, More Fundamental Case for Social Media Agent 'Middleware,'" Smartly Intertwingled, November 9, 2023, https://ucm.teleshuttle.com/2023/11/a-new-broader-more-fundamental-case-for.html.

[61] Benjamin Lauffer and Helen Nissenbaum, "Algorithmic Displacement of Social Trust," Knight First Amendment Institute at Columbia University (November 29, 2023), https://knightcolumbia.org/content/algorithmic-displacement-of-social-trust.

The ability to determine credibility and authoritativeness is essential for social cohesion and foundational to democratic governance. In response to this challenge, social media platforms have often defaulted to upleveling content backed by institutional expertise, though the COVID-19 pandemic highlighted flaws in this approach. The institutions posted far less frequently, for example, than small accounts belonging to frontline doctors working in emergency rooms. These smaller accounts often provided highly authoritative insights, but were hard to find amid the deluge of content. However, the challenge is compounded by the fact that anyone can present as a frontline doctor by simply creating an account that claims that they are; it is difficult for users to determine whether someone is accurately representing themselves, and whether they are trustworthy.[62] Systems such as "blue check" verified accounts were originally designed to create signals of trustworthiness, but not all platforms award such credentials in the same way.

Middleware products might also be used to surface indicators of reputation and authoritativeness on current social media platforms, and incorporate users into making those determinations, thereby increasing trust. As a simple example, consider a middleware product similar to the now-defunct Klout, which assigned topical expertise labels and authoritativeness scores derived from user content assessed by Klout itself, as well as community "vouches" in which users awarded each other "+K" in specific areas of expertise.[63] By leveraging social associations and community-ascribed expertise, middleware could foster greater trust in information and institutions in ways that promote bridging community divides.[64]

The idea of a "reputation layer" for the internet is not new, and there are providers who have begun to leverage various technologies to create it. Reputation presumes and builds on a reliably persistent identity.[65] A persistent pseudonymous identity middleware provider might be useful for creating a portable reputation system akin to Reddit's Karma score. A bluecheck confirming that a user is the person they claim to be might support trust across the entirety of the social media ecosystem, offering a credential independent of the whims of any one large platform provider. This system could involve multi-factor "vouchers" that could confirm either identity, credentials, or expertise collaboratively, making it easier for users to identify trustworthy sources. A professional credentialing organization, such as a medical board, might contribute to a tool to surface its members, increasing trustworthiness in context-specific areas.

---

[62] Renee DiResta, *Invisible Rulers: The People Who Turn Lies Into Reality* (New York Hatchette, 2024), Chapter 7.
[63] Joel Falconer, "Klout Introduces +K Peer Verification for Social Influence," *The Next Web*, June 2, 2011, https://thenextweb.com/news/klout-introduces-k-peer-verification-for-social-influence.
[64] Aviv Ovadya and Luke Thorburn, "Bridging Systems: Open Problems for Countering Destructive Divisiveness Across Ranking, Recommenders, and Governance," Knight First Amendment Institute at Columbia University (October 26, 2023), https://knightcolumbia.org/content/bridging-systems.
[65] Many projects related to "Web3" or the decentralized web are exploring decentralized identity and reputation systems, leveraging blockchain technology in attempts to create persistent, portable identities that are not controlled by any single platform. For example, projects such as the Ethereum Name Service and Ceramic Network are building decentralized identity frameworks that could enable users to maintain a verifiable reputation across multiple platforms. Additionally, decentralized social media projects such as Lens Protocol are experimenting with portable, user-controlled reputation systems and social graphs that allow individuals to build trust across different ecosystems without relying on centralized intermediaries.

Similarly, for authoritativeness, fact-checking organizations could create subscribable feeds; more broadly, middleware could support collaborative fact-checking tools, akin to X's Community Notes, that allow users to flag and verify content collectively and transparently. These tools could integrate with reputation systems, where users known for accuracy and reliability in fact-checking are given greater weight in the process.

Addressing the deeper issues of reputation, authenticity, authority, and credibility is necessary to creating a more transparent and trustworthy online environment, and middleware may offer some novel paths forward.

## Concerns with Middleware

Middleware is not a panacea that will solve all of the problems of the internet; there are trade-offs, particularly when it comes to the specifics of implementation. While centralized platforms with opaque curation and moderation algorithms may spark fears of digital tyranny, fully devolved decentralization and user control of moderation and curation spark fears of chaos. Middleware is fundamentally an enabling technology, and finding the proper balance of top-down and bottom-up control for any given challenge will be essential to creating a healthy and inclusive internet.

Several criticisms of middleware will be considered at length in this section, including concerns about accelerating or exacerbating echo chambers, potential gaps in moderation capacity, the need to ensure adequate privacy protections for middleware users, and questions around the technical feasibility of middleware ("is it even possible to do this?").[66]

First, we consider the most common criticism levied against middleware: that it trades the tyranny of centrally controlled platforms for the chaos and social fragmentation of user-controlled ultra-individualized experiences. Does middleware optimize for divisiveness, rather than collective decision-making? Some have argued that middleware could foster further polarization or fragmentation of society.[67] Where curation is concerned, enabling users to self-select their preferred sources of content, or to opt out of a centralized platform experience, might *reduce* their exposure to diverse viewpoints, leading to more siloed, bespoke online realities. The desire to select feeds aligned with personal beliefs or emotional preferences could amplify confirmation bias, facilitate or reinforce belief in false information, or reduce engagement with challenging or alternative perspectives. In extreme cases, middleware might facilitate the expansion or strengthening of deeply conspiratorial communities (consider the possibility of the "QAnon feed"). Unlike centralized systems where labeling efforts for false or misleading content reach all users, middleware would rely on opt-in mechanisms, which could be problematic given that trust in fact-checkers is often divided along partisan lines.[68] Over time, such fragmentation could deepen political and cultural divides, erode the shared informational foundation essential for democratic debate, and hinder collective action or consensus-building.

---

[66] Daphne Keller, "The Future of Platform Power: Making Middleware Work," *Journal of Democracy,* vol. 32, number 3 (July 2021), https://muse.jhu.edu/article/797795.
[67] "Francis Fukuyama: Middleware Could Give Consumers Choices Over What They See Online," *Stanford HAI,* November 17, 2021, https://www.youtube.com/watch?v=zrfIRfJxwQM.

There are similarly tyranny-versus-chaos tradeoffs in the realm of moderation middleware as well. Customizable moderation feeds on largely decentralized, federated platforms—where no top-level moderation exists for explicitly illegal or violent content—introduces significant challenges for content moderation and risks creating negative externalities for other users.[69] In the case of centralized platforms, even those that incorporate middleware will likely continue to moderate certain classes of content, such as that which is illegal, violent, or threatening, at the top. Even in the case of content that does not rise to direct incitement to violence, the consequences of groups of users deliberately, for example, mobbing an individual with slurs, threats, or vicious images have business implications: it can lead to users feeling chilled, decreasing overall participation.[70] Policies around lawful-but-awful content on centralized platforms therefore generally attempt to create environments in which disrespectful and harassing content is minimized or discouraged to create a positive user experience for the participants and a brand-safe experience for business.

On a decentralized but interconnected system, that incentive is diffused, and sometimes it is simply technologically impossible for a decentralized instance to act—it cannot control what is shared to other servers, so doxing and harassing content, for example, may reach the community from elsewhere. Middleware that hides harmful content from the target's view does little to mitigate the harm if the content remains visible to others. Even explicitly illegal and violent content has a far higher likelihood of remaining up on decentralized sites whether middleware is present or not.[71]

---

[68] Cameron Martel and David Rand, "Fact-Checker Warning Labels are Effective Even for Those Who Distrust Fact-Checkers," *Nature Human Behaviour,* issue 8, September 2, 2024, https://www.nature.com/articles/s41562-024-01973-x.
[69] Yoel Roth and Samantha Lai, "Securing Federated Platforms: Collective Risks and Responses," *Journal of Online Trust & Safety*, vol. 2, no. 2 (2024), https://tsjournal.org/index.php/jots/article/view/171.
[70] Chris Bail, *Breaking the Social Media Prism: How to Make Our Platforms Less Polarizing* (Princeton: Princeton University Press, 2021).
[71] David Thiel and Renee DiResta, "Addressing Child Exploitation on Federated Social Media," *Stanford Internet Observatory,* July 24, 2023, https://cyber.fsi.stanford.edu/io/news/addressing-child-exploitation-federated-social-media.

User-selected moderation middleware also risks fragmenting community standards, making it hard to establish and maintain shared norms. Some users might choose tools that permit or even encourage harmful behavior, leading to potential abuse or discrimination—particularly for marginalized or minority communities of users. Conversely, middleware moderation tools might be overly stringent: shared blocklists with vague or opaque criteria can create unintended consequences, or might be used to target users personally disfavored by a creator or administrator. Users may fear being added to a blocklist for expressing dissenting views, leading to self-censorship and a chilling effect on open dialogue.

It is also unclear what authority, if any, will be responsible for monitoring middleware to assess security, privacy, and to avoid perpetuating discriminatory policies. These issues illustrate the tension inherent in middleware: while it offers users greater agency, it can also amplify fragmentation, creating environments that vary widely in terms of safety, inclusivity, and trustworthiness. While middleware might empower users to choose services that better align with their values, or privacy and security preferences, it also raises new risks. If providers are not held to strict standards, we risk replicating or even exacerbating existing issues with platform data collection or poor content moderation.[72]

Indeed, one major challenge that middleware presents is how to build open and interoperable networks while simultaneously protecting user privacy. It is simple enough to allow a user to give a middleware provider consent to read and organize ones' private data, but it is a much thornier question to consider what a middleware provider's relationship should be with a user's friend's privately shared posts or other information, especially when that friend never consented to having their information shared with that third party. In this instance, protecting user privacy would mean limiting middleware's access to certain types of content, especially private posts. However, this could hinder the very goals middleware is designed to achieve, such as giving users greater control over how content is curated. Addressing these privacy issues will require balancing user privacy with interoperability and content moderation. Regulatory solutions that address how middleware interacts with both user data and friends' data will be crucial to making middleware effective without compromising privacy.

While some may assume that user privacy can only be protected by a centralized controller, there are various models available for establishing trust when third-party developers access sensitive data. One of the most prevalent models, of course, is monolithic control by platforms, exemplified by mobile app stores where companies like Apple and Google set strict standards, unilaterally evaluate the trustworthiness of third parties' privacy practices, and maintain sole authority to authorize apps. Alternatively, crowd-sourced community standards, such as

---

[72] Daphne Keller, "The Future of Platform Power: Making Middleware Work," *Journal of Democracy*, vol. 32, issue 3 (July 2021), https://www.journalofdemocracy.org/articles/the-future-of-platform-power-making-middleware-work/.

Spamhaus or Project Honeypot, rely on collaborative input to create norms that platforms may voluntarily adopt. Another approach involves industry standards and formal certification processes, such as the WebTrust/CA Browser Forum or TrustArc, which can include independent auditing to ensure compliance. By considering these frameworks, policymakers and developers can envision middleware ecosystems that balance innovation, accountability, and user protection without defaulting to unilateral platform control.

While middleware offers promising solutions to many challenges posed by centralized platforms, it is not immune to the economic pressures that have shaped the current social media landscape. If the middleware market becomes a significant "source of eyeballs," providers might compete for user adoption in ways that replicate many of the problems on existing platforms. It is possible that most users will not actively seek out middleware providers to serve them, leaving dominant players to emerge based on their ability to achieve the lowest cost of acquisition, the most addictive features, and the highest revenue per customer—factors that could drive rapid revenue growth but also perpetuate problematic dynamics.

This suite of concern will take time to sort out—in practice, technology, and policy. There is an argument that some top-down control is needed, with a question of how that blends downward community control and more formal government control (as addressed further in the policy section).[73] One possible approach is a two-tier middleware structure: a tightly controlled layer for sensitive core functions, potentially with formal certification, supporting a more open layer that allows for greater diversity in user-agent functionalities.[74] Over time, hybrid levels of community control might mature to require less top-down action. A key advantage of middleware, however, is that it empowers users to decide how much community influence they wish to adhere to, and, based on those preferences, to choose the communities with which they engage.

To navigate these challenges, it is crucial to recognize the dual role of middleware as both an enabler of user choice and a potential amplifier of social dynamics. By establishing frameworks that evaluate middleware's impact on bridging or dividing communities, stakeholders can better align its development with broader societal goals.

Consider the specific risk of middleware entrenching polarization. It is theoretically possible to create a suite of metrics to evaluate middleware, focusing on factors such as the content being shown, the algorithms themselves, or the interfaces provided. In the case of polarization, these might assess the extent to which a middleware system fosters bridging behaviors—ranking content that connects users across divides—or whether it amplifies polarizing dynamics.

---

[73] See Yoel Roth and Samantha Lai, "Securing Federated Platforms: Collective Risks and Responses".
[74] For example, the privacy of content, flow, and behavior metadata that can be used to support a loosely managed and very diverse tier of middleware ranking services that are both powerful and open might be assured using a smaller, tightly managed tier of data intermediaries, as suggested by Richard Reisman, "How Third-Party Social Media Middleware Can Protect Contextual Privacy," *Tech Policy Press*, October 11, 2023, https://www.techpolicy.press/how-third-party-social-media-middleware-can-protect-contextual-privacy/. See also Jonathan Stray, "A Concrete Proposal for a Middleware API," April 30, 2024, https://docs.google.com/presentation/d/1B4jCnHu6Ggr0_ODDzuuLW4oX8lYim61-qfqY9qGKXxM/edit#slide=id.gc8f20ba54b_0_0 .

Using these metrics, policymakers or standards bodies could implement various mechanisms to encourage positive development of middleware. Given such bridging metrics around polarization, it is possible to create defaults, incentives, warnings, or even requirements around the metrics (e.g., via a standards and certification body). Middleware providers could be required to meet minimum thresholds on bridging metrics or have specific certifications. Incentive programs might tie platform revenue-sharing agreements to higher performance on these metrics, rewarding providers that promote healthier online environments. There might also be requirements to warn users if they choose middleware providers with low metrics. At the extreme, middleware providers with persistently poor performance could face decertification and be excluded from use on certain platforms.

This approach of defaults, warnings, incentives, and floors oriented around metrics can apply to many middleware applications, and potentially help shift the competitive landscape from a race to capture attention at any cost to one that prioritizes quality and accountability. There are significant challenges in developing such metrics,[75] but there has also been significant progress recently in collecting information about such pro-societal metrics (in addition to work around measuring the societal harms).[76] Continuing to invest in this line of work can help to address the key risks of middleware and to achieve its promise.

One primary challenge with such approaches is that there remains the question of "who decides" what these metrics, certifications, defaults, requirements, and so on are. Protocol standard bodies are likely to play a significant role in the technical specifications, but they are perhaps less well suited for making decisions around societal metrics and certification requirements. This process may instead involve industry consortia and multi-stakeholder bodies, and for the most sensitive and controversial questions where there is "no good decider" (e.g., around baseline requirements for certification), deliberative processes through platform democracy may prove useful.[77]

---

[75] Tom Cunningham, et al., "What We Know About Using Non-Engagement Signals in Content Ranking," Arxiv, February 9, 2024, https://arxiv.org/abs/2402.06831.
[76] Justin Hendrix and Jonathan Stray, "Ranking Content On Signals Other Than User Engagement," *Tech Policy Press*, February 18, 2024, https://www.techpolicy.press/ranking-content-on-signals-other-than-user-engagement/.
[77] Aviv Ovadya, "Towards Platform Democracy: Policymaking Beyond Corporate CEOs and Partisan Pressure," Belfer Center for Science and International Affairs, October 18, 2021, https://www.belfercenter.org/publication/towards-platform-democracy-policymaking-beyond-corporate-ceos-and-partisan-pressure.

Technical Feasibility of Middleware Solutions

Having made a case for the potential of middleware to enhance user and community choice, and protect democratic freedoms—and having considered possible downsides and negative externalities—we must also consider the question of how difficult it might be to implement. Here we discuss some of the technical challenges of responsible, secure middleware. In its fullest conception, open, third-party middleware between platforms and users is technically challenging to implement. However, there are simpler approaches and solutions that can support more limited middleware capabilities. Thus, technical feasibility should be viewed as a matter of ongoing development and maturation of both the technology and how it is used.

For example, consider a small startup that wishes to create a piece of middleware for X that allows users to completely replace the default content ranking algorithm with a personalized curation alternative. There are perhaps 500 million items posted on X each day,[78] meaning that any ranking algorithm must select from at least hundreds of millions of posts even if only the most recent material is considered. The platform must therefore make this material available to a would-be middleware provider. The most straightforward solution is for X to offer a data access API whereby the middleware provider can continually download all of this material from the platform's servers.

This is a phenomenal volume of material—merely ingesting this much data in real time is a challenging engineering feat, and the network and storage costs are not insignificant. A content ranking service must consider all of this material in a fraction of a second when a user opens the app. To achieve this, platforms maintain a variety of indices, continually cataloging and clustering content based on topic, engagement patterns, metadata, and more, which a middleware provider would have to duplicate, or use specialized shared services to accomplish, as explained below for the example of Bluesky. This level of engineering is well beyond what a garage-sized startup (or small middleware provider) can afford unaided.

However, it may not be necessary for every middleware service to independently ingest and process all of the data. One way to potentially minimize the computational and storage requirements for middleware providers might be to have an intermediary set of services that do most of the heavy lifting, which user-facing middleware services are then built on top of. For example, some platforms like Bluesky provide for independent indexers, so that a small number of independent, secure data intermediary services that are resource intensive can serve many lighter-weight Bluesky middleware feed composing services, to work with any of multiple client apps, all using the AT Protocol.[79] Similarly, Bluesky is working toward a "stackable approach to moderation," that allows users to create a moderation layer on top of the base Bluesky app

---

[78] "Twitter Usage Statistics," Internet Live Statistics, retrieved November 4, 2024, https://www.internetlivestats.com/twitter-statistics/.
[79] Kleppmann et al., "Bluesky and the AT Protocol."

"without running a lot of infrastructure or building your own client app."[80] Moderation, e.g. labeling, obscuring, or removing unwanted content, does not require processing all content on the platform, but merely filtering the small volume of content already selected for a particular user. Implementing a global content ranking system, which might process every item on a platform, is a much more challenging problem. Scale problems are far more pressing when considering middleware for video platforms like YouTube where data volumes are orders of magnitude larger than for text and still images.

In any case, as noted above, there are serious privacy concerns with middleware that must be considered and addressed during technical development. While posts on sites like X, Bluesky, and Threads are public by default, posts on other platforms like Facebook are private by default and can only be seen by a users' friends. If a Facebook user installs middleware, it may be the case that a friends' posts are transferred to the middleware provider so that they can be ranked and displayed. While a middleware user might consent to sharing their private data with a middleware provider, without proper guardrails, data may also be transferred from users who never consented—and who may not even be aware that their data is being shared with a third party. Such potential privacy issues are thorny but may be surmountable with a combination of technical approaches and consumer protection regulations.

Both scale and privacy considerations argue for a more efficiently structured approach: instead of moving platform data to a middleware provider's servers, one could move the middleware provider's software to platform servers. In this model, the middleware software would make use of platform or secure third-party service resources including server hardware, content databases, and whatever indices or pre-processed data already exist. This would require a much more complex platform API. Instead of mere data access, the platform or intermediary would have to offer the ability to deploy third-party software on its own servers,[81] possibly along with associated services such as the ability to provision additional computation, custom indices, storage, and so on.

This model solves two key problems: it could drastically reduce the startup cost for middleware market entrants, and, for non-public messaging, virtually eliminate the privacy and security concerns that arise from transferring large volumes of private user data to third parties. In fact, middleware operators need never see any user data at all—they would throw their code "over the wall" into a platform sandbox.

---

[80] "Bluesky's Stackable Approach to Moderation," Bluesky, March 12, 2024, https://bsky.social/about/blog/03-12-2024-stackable-moderation.
[81] This idea has been mentioned before. See, e.g., Stephen Wolfram, "Testifying at the Senate about A.I.-Selected Content on the Internet," Stephen Wolfram Writings, June 25, 2019, https://writings.stephenwolfram.com/2019/06/testifying-at-the-senate-about-a-i-selected-content-on-the-internet/.

Although this may seem like a tidy solution, it does ask a lot from platforms and could be expensive for them to implement. In effect, the platform or intermediary would be transformed into a specialized cloud services provider, closer to Amazon Web Services or Google Cloud than a mere API. It is unlikely that platforms would invest the required resources without either a profitable payment/revenue-share model or regulatory requirements.

Fortunately, there are simpler technical approaches to middleware which do not provide the ability to create full replacements for platform ranking systems, yet would still enable useful third party capabilities for middleware in content curation.

First, a middleware provider could select content from a much smaller set of user accounts, or only from items which match narrow predefined filters, rather than attempting to ingest and potentially select from among every item on the platform. This is the typical approach of federated systems like Mastodon, where current APIs are more oriented around creating "filters" or custom content moderation tools than general content rankers or recommenders.[82] In any case, the content volume on these platforms is currently orders of magnitudes smaller than on large commercial platforms, as they have far fewer users. While platforms do not often rank and recommend content from smaller sets of user accounts as described above, such systems are possible and could add value for users.

Second, middleware could operate by inserting itself into the stream of content sent from the platform to the client app. This would allow reordering, adding, or removing items from each batch of posts sent to the user. While this is not a fully general middleware solution, it is perhaps the easiest to implement, both for platforms and middleware creators. This technique is already used by academic researchers to test alternative social media algorithm designs without platform permission.[83]

# **Markets for Middleware Development**

While middleware offers significant potential to address the challenges of centralized platforms, the concerns outlined above highlight the complexity of building and sustaining a middleware ecosystem. Middleware will remain niche until there are markets to facilitate it, and to make it a scalable and potentially profitable (or otherwise worthwhile) endeavor for the providers. As the Stanford Working Group on Platform Scale acknowledged in its original paper, market incentives must be aligned for middleware to flourish, and "a business model for middleware providers must be sufficiently attractive to induce an adequate supply."[84] In this section, we analyze what a market for middleware development might look like, discuss how to align market incentives to maximize the potential of middleware, and offer suggestions for how this market might evolve in either cooperative or adversarial ways.

---

[82] See "Bluesky's Stackable Approach to Moderation," Bluesky.
[83] See, e.g. Piccardi et al., "Reranking Social Media Feeds: A Practical Guide for Field Experiments," June 27, 2024, https://arxiv.org/abs/2406.19571, and Stray, "The Prosocial Ranking Challenge," Center for Human-Compatible Artificial Intelligence, January 18, 2024, https://humancompatible.ai/news/2024/01/18/the-prosocial-ranking-challenge-60000-in-prizes-for-better-social-media-algorithms/.

## Means of Adoption

Given the current dominance of large, centralized social media platforms, strategic alignment between incumbents and upstart middleware services is an important driver for galvanizing the development of middleware. There are two routes that create momentum toward increased openness: platform-driven or user-driven.

The most direct route to creating a vibrant market for middleware would be a scenario in which incumbent platforms are incentivized (or required) to open themselves up. To some degree, we are already seeing tentative moves toward platform-driven adoption of middleware. In the last few years, several of the largest social media companies and platforms have taken steps towards more openness and interoperability. Then-CEO of Twitter Jack Dorsey originally funded Bluesky with the intent of creating an open protocol-based middleware ecosystem for Twitter; BlueSky has since been spun out as an independent entity. More recently, Dorsey has backed Nostr as an independent, open protocol-based architecture for Twitter-like social media.[85] Meta has also moved toward enabling middleware through its recent move to integrate ActivityPub—a decentralized, open social networking protocol standardized through the World Wide Web Consortium—to connect its new Threads app (a Twitter-style product) with the fediverse of Mastodon and other compatible services.[86]

Of course, platforms usually do not open themselves up for purely altruistic reasons; companies must see some benefit to openness. With regard to Meta's integration of ActivityPub, the company likely sees a market incentive in the possibility of onboarding and connecting with more users from the fediverse. There is also perhaps a policy incentive pushing it towards openness: regulation and potential legal action from the European Union, discussed in greater detail later in this paper in the section on public policy.[87] This "carrot-and-stick" incentive structure can help lead to a platform-driven adoption process.

One example of market incentives driving adoption of more open systems that allow third-party software is app stores. When the iPhone launched in 2007, it had only about a dozen apps, all developed by or under contract by Apple; Steve Jobs did not initially intend to open that ecosystem.[88] However, the company quickly saw that there was high demand for a diversity of third-party applications. In choosing to open up its App Store, Apple enabled a market of independent software and service providers under relatively limited platform control. Apple now boasts that its App Store ecosystem generates $1.1 trillion in annual billing and sales.[89] Perhaps more importantly for Apple, the ability to use a diverse set of third-party apps has increased the utility, and thus demand for, iPhones.

---

[84] See Fukuyama et al., "Report of the Working Group on Platform Scale."
[85] Katherine Long, "Jack Dorsey Gave $10 million to an Anonymous Founder with a Deep Devotion to a Fascist 'Guru,'" *Business Insider,* June 6, 2024, https://www.businessinsider.com/jack-dorsey-fiatjaf-nostr-donation-2024-6.
[86] Christopher Su and Simon Blackstein, "Threads Has Entered the Fediverse," Meta, March 21, 2024, https://engineering.fb.com/2024/03/21/networking-traffic/threads-has-entered-the-fediverse/.
[87] Luke Hogg and Lars Eric Schönander, "Meta's Interoperable Reversal," *Tech Policy Press,* July 19, 2023, https://www.techpolicy.press/metas-interoperable-reversal/.

As this App Store analogy suggests, financial incentives can encourage openness, leading to a thriving and resilient market that supports innovative product ecosystems that non-technical users can use with ease and proficiency. More recently, however, Apple has faced criticism around its handling of the App Store; critics argue that Apple uses its control over the App Store to stifle competition and impose artificially high take rates on mobile developers.[90] In response, the European Union has mandated that Apple allow third-party app stores on iOS, and several lawmakers are proposing that the United States implement similar requirements. Sustaining platform openness often involves both carrots and sticks.

Instead of platforms driving the emergence of middleware, a market could alternately emerge through user demand, as newer social media platforms demonstrate its potential. For example, as Bluesky's popularity has skyrocketed throughout November and December of 2024—as of the date of publication, it has reached 20 million users[91]—Meta's Threads, which had already incorporated the ActivityPub protocol and begun to integrate with the fediverse, has also announced that it will launch features such as user-created curation feeds that Bluesky users have praised.

A user-driven approach to creating a market for middleware has traditionally been envisioned as far more adversarial, relying on a "build it and they will come" ethos based on disruption via adversarial interoperability—the practice of creating products or services that can interoperate with existing systems or platforms without the cooperation or permission of the original provider.[92] This approach allows smaller, more agile players to innovate rapidly and potentially achieve critical mass network effects—where the value of product or service is relative to the number of people using it—without waiting for large platforms to open up (although it may also expose them to liability if platforms sue). This strategy can foster a diverse range of middleware tailored to niche needs, but has significant challenges. Even with the rise of federated platforms, new entrants face high barriers to entry, including the costs of scaling operations and competing with established giants.

---

[88] Stuart Dredge, "Steve Jobs Resisted Third-Party Apps on iPhone, Biography Reveals," *The Guardian*, October 24, 2011, https://www.theguardian.com/technology/appsblog/2011/oct/24/steve-jobs-apps-iphone.
[89] "App Store Developers Generated $1.1 Trillion in Total Billings and Sales in the App Store Ecosystem in 2022," Apple, May 31, 2023, https://www.apple.com/cm/newsroom/2023/05/one-point-one-trillion-generated-in-app-store-ecosystem-in-2022/.
[90] Sarah Needleman and Aaron Tilley, "Apple Changes Its App Store Policy. Critics Call the Moves 'Outrageous,'" *Wall Street Journal,* January 17, 2024, https://www.wsj.com/tech/apple-changes-its-app-store-policy-critics-call-the-moves-outrageous-7c023e0c.
[91] Bluesky Post Count and Author Stats, accessed December 3, 2024, https://bsky.jazco.dev/stats.
[92] Cory Doctorow, "Adversarial Interoperability: Reviving an Elegant Weapon From a More Civilized Age to Slay Today's Monopolies," Electronic Frontier Foundation*,* June 7, 2019, https://www.eff.org/deeplinks/2019/06/adversarial-interoperability-reviving-elegant-weapon-more-civilized-age-slay .

Absent a significant event—such as a major change in platform policy that leaves users dissatisfied and actively seeking alternatives[93]—new platforms and middleware providers must attract users away from entrenched systems by investing in marketing, user education, and creating features that offer clear benefits over existing options. Bluesky, for example, is finding an impressive level of success with this go-it-alone strategy, capitalizing on its appeal to emigres from X.

Potential Business Models

There is growing demand from users to enjoy greater safety, quality, and personalization in their social media experience. Prior to the rise of entirely new options, such as Bluesky, this was evidenced by the adoptions of tools such as the popular anti-harassment provider Block Party. However, in order to expand, scale, and innovate, finding sustainable and legal business models for potential providers is key.

Early social media services used a wide variety of business models to fund their operations, including volunteer self-funding, micropayments, and subscriptions. As social media reached mass scale in the 2000s, large centralized platforms came to dominate, driven by economies of scale that lowered operational costs, and network effects that locked in users. Large user bases lent themselves to an advertising business model; however, this required that users be on the platform to see the ads, and necessitated significant data gathering to ensure precision targeting. This approach, sometimes called "the internet's original sin,"[94] incentivized maximizing users' time on site, regardless of whether that time was "well spent." It generated enormous profits, but discouraged platforms from empowering users with greater agency over their feeds and recommenders, since, at least in the short run, that might reduce engagement and revenue per user.[95]

---

[93] Alex Hern, "Twitter Usage in US 'Fallen by a Fifth' since Elon Musk's Takeover," *The Guardian*, March 26, 2024, https://www.theguardian.com/technology/2024/mar/26/twitter-usage-in-us-fallen-by-a-fifth-since-elon-musks-takeover.
[94] Ethan Zuckerman, "The Internet's Original Sin," *The Atlantic*, August 14, 2014, https://www.theatlantic.com/technology/archive/2014/08/advertising-is-the-internets-original-sin/376041/.
[95] Georgia Wells, Jeff Horwitz, and Deepa Seetharaman, "Facebook Knows Instagram Is Toxic for Teen Girls, Company Documents Show," *Wall Street Journal,* September 14, 2021, https://www.wsj.com/articles/facebook-knows-instagram-is-toxic-for-teen-girls-company-documents-show-11631620739.

Today, most large platforms still rely on advertising as their primary revenue source, while many smaller social media services struggle to sustain themselves on ads alone. One key challenge for middleware markets on large platforms is determining how middleware can operate without directly competing for, or solely relying on, ad revenue. The following sections explore potential middleware revenue models, including for-profit, non-profit, and hybrid approaches.

### For-Profit Models

A direct ad model for middleware services is possible but may conflict with platform ad models. Ad revenue share models could be negotiated between large platforms and middleware providers, much as is done for cable TV ads, website syndication services, and many platforms that rely on user-generated content.[96] For example, Meta could integrate middleware services into its current "Audience Network" of third-party apps, which helps both Facebook and third parties generate ad revenue in a symbiotic way.[97]

Subscriptions offer another revenue model for both social media communities and for middleware providers. Some large platforms are diversifying their revenue streams by incorporating subscriptions; for example, after purchasing Twitter (now X), Elon Musk took steps toward a subscription-based "freemium" model, providing basic services for free while upselling users on premium subscription features (such as preferential placement in curation).[98] Meta, TikTok, and Snapchat are also reportedly testing or planning subscription models.[99] Some middleware services could generate subscription revenue from users who value higher quality curation, feel exposed to objectionable content, or are concerned about privacy. Indeed, Block Party operated under a "freemium" model before being forced to shut down the tool after losing access to X's API.[100] There is also the possibility of more creative arrangements where platforms might offer third-party middleware to their users as premium products with a negotiated revenue share.

Middleware providers could also rely on other kinds of user service fees and micropayments to generate revenue. This model is being explored by services such as Bluesky, which now takes a cut from the creation of individual web domains to be used as persistent, cross-service identifiers on the platform.[101] Many blockchain-based services use cryptocurrency micropayments per unit of service to support the maintenance of the network, as the Filecoin ecosystem does for digital storage.[102] Some forms of service fees and per-unit micropayments have been reasonably successful for some online services, but they tend to encounter a problem where users fear the "ticking meter" and surprisingly large bills.[103]

---

[96] Francis Fukuyama, Barak Richman, and Ashish Goel, "How to Save Democracy From Technology: Ending Big Tech's Information Monopoly," *Foreign Affairs*, November 24, 2020, https://foreignaffairs.com/articles/united-states/2020-11-24/fukuyama-how-save-democracy-technology.
[97] "Meta Audience Network," Meta, accessed October 8, 2024, https://www.facebook.com/audiencenetwork/.
[98] Arjun Kharpal, "X, Formerly Twitter, Will Launch Two New Subscription Tiers, Elon Musk says," CNBC, October 20, 2023, https://www.cnbc.com/2023/10/20/elon-musk-x-formerly-twitter-to-launch-two-new-subscription-tiers.html.
[99] Daniel Avshalom, "Advertisers, Brace Yourselves: The Era Of Paid, Ad-Free Social Media Is Here," AdExchanger, October 23, 2024, https://live-adexchanger.pantheonsite.io/data-driven-thinking/advertisers-brace-yourselves-the-era-of-paid-ad-free-social-media-is-here/.
[100] Sarah Perez, "After Losing Access to Twitter's API, Block Party Pivots to Privacy," *TechCrunch,* March 11, 2024, https://techcrunch.com/2024/03/11/after-losing-access-to-twitters-api-block-party-pivots-to-privacy/.

Another approach that is applicable to middleware is revenue sharing or even reverse metering.[104] Much as electric meters run in reverse when a solar panel feeds electricity back into the grid, online service payments can run in reverse when users provide value back to the service in the form of ad views, data for targeting ads, or in other forms such as user-generated content. Middleware providers might use similar reverse metering to encourage users to provide community notes and other forms of explicit feedback that can be valuable in curation and moderation, thus offsetting possible subscription fees. A unifying model for adaptively applying such diverse kinds of fees as subscriptions, micropayments and reverse metering is provided by the FairPay framework[105]

## Nonprofit Models

Technology does not always need to be a for-profit endeavor. Platforms can be funded in the public interest, whether by organizers, users, government, or philanthropists. If middleware is seen as a socially valuable good, providers could operate as nonprofits, as many open-source projects already do. There are essentially two basic questions that must be answered when considering nonprofit models for middleware development: what sort of governance will be utilized, and where will funding come from?

---

[101] Bluesky Team, "Purchase and Manage Domains Directly Through Bluesky, Bluesky, July 5, 2023, https://bsky.social/about/blog/7-05-2023-namecheap.
[102] "Crypto-Economics," Filecoin Docs, accessed October 8, 2024, https://docs.filecoin.io/basics/what-is-filecoin/crypto-economics.
[103] Richard Reisman, "'The Case Against Micropayments' versus 'Subscription Hell' -- Finding Flexibility," *FairPay Zone,* November 13, 2018, https://www.fairpayzone.com/2018/11/the-case-against-micropayments-from.html.
[104] Jeff Jarvis, "Why Not a Reverse Meter?" *BuzzMachine*, December 19, 2011, https://buzzmachine.com/2011/12/19/why-not-a-reverse-meter/.
[105] Richard Reisman, "The Elements of Next-Gen Relationships and Pricing -- A Unifying Framework," *FairPay Zone*, July 3, 2019, https://www.fairpayzone.com/2019/07/the-elements-of-next-gen-relationships.html. These dynamic models may themselves become an application for middleware services that serve as user agents to negotiate pricing for middleware and other social media services. This framework is based on the models detailed by Richard Reisman, Adrian Payne, and Pennie Frow, "Pricing in Consumer Digital Markets: A Dynamic Framework," Australasian Marketing Journal, vol. 27, issue 3 (August 1, 2019), https://journals.sagepub.com/doi/10.1016/j.ausmj.2019.07.002, which references studies of a variety of innovative revenue models that deserve consideration for middleware services.

When it comes to nonprofit governance, middleware providers could adopt traditional hierarchical structure, though alternative options also exist. Organizational structures that seek to provide beneficial services include cooperatives, which are owned by and serve their members, and benefit corporations, which can earn a profit, but are legally required to prioritize goals beyond profit alone; Bluesky, for example, operates as a benefit corporation.[106] There is also a movement seeking to develop "platform cooperatives" or "data cooperatives,"[107] as well as interest in decentralized autonomous organizations (DAOs), a still-newer organizational structure for managing sharing of user data, typically based on blockchain technology and using cryptocurrency for governance.[108]

Starting a middleware project, even with the intent of a non-profit model, requires significant startup funding; for-profit models are often initially funded by the entrepreneurs themselves, or via friends and family, venture capital, or angel investors. When it comes to securing funding for a nonprofit middleware service, there are various avenues. Mastodon, for example, is open-source software developed by a non-profit organization; it accepts donations.[109] Many nonprofits are funded purely through philanthropy, and there are many foundations and wealthy individuals who provide grants to nonprofit projects they see as aligned with their values and goals. Whether as a form of capital investment or philanthropy, funding by billionaires has become a significant factor in the emergence of middleware-based online services. Jack Dorsey was a major source of early funding for both Bluesky and Nostr, and Project Liberty's Decentralized Social Networking Protocol is backed by billionaire Frank McCourt. This funding calls attention to obvious risks of improper oligarchic influence over discourse, but can be a way to jump-start a service that is organized to be independently controlled, as with Bluesky and Nostr.

Some experimental social media services, such as Gobo and other projects of the Initiative for Digital Public Infrastructure at UMass Amherst, are funded by grants. As the potential social value of middleware becomes better recognized, this could be a significant source of sustaining funds, especially for early-stage development. Government grants are an important source of funding for public interest infrastructure and technologies of all kinds. Middleware providers seeking to use a purely nonprofit model should consider casting a wide net in search of funding, as any one of these sources may not be sufficient to fund a project in perpetuity.

---

[106] "Our Plan for a Sustainably Open Social Network," Bluesky, July 5, 2023, https://bsky.social/about/blog/7-05-2023-business-plan
[107] Katharine Miller, "Radical Proposal: Data Cooperatives Could Give Us More Power Over Our Data," Stanford HAI, October 20, 2021, https://hai.stanford.edu/news/radical-proposal-data-cooperatives-could-give-us-more-power-over-our-data.
[108] Sarah Hubbard et al., "Toward Equitable Ownership and Governance in the Digital Public Sphere," Belfer Center for Science and International Affairs, Harvard Kennedy Center, June 8, 2023, https://www.belfercenter.org/publication/toward-equitable-ownership-and-governance-digital-public-sphere.
[109] Mastodon, "Mastodon Annual Report 2022," accessed December 1, 2024, https://joinmastodon.org/reports/Mastodon%20Annual%20Report%202022.pdf.

Many online community services—particularly open-source projects such as Linux, Tor, Wikipedia, and Apache—rely heavily on volunteer labor. This can be useful at small scale, or in large networks composed of small-scale services, as in the Mastodon fediverse, but generally presents unique challenges as scale increases. Building open-source tools in a modular way where future developers can "plug and play" or benefit from Software as a Service (SaaS) offerings can reduce some of the development and labor costs and other challenges that scale brings. Similarly, users can make voluntary payments or participate at varying levels in making contributions to operating costs. This is a time-tested way to make services sustainable, often with some variation on pay-what-you-want, pay-what-you-can, tipping, and donationware. Conventional uses of such methods have been difficult to scale with good results, but more innovative variations may emerge as user demand increases.

## Future Market Opportunities

The phenomenon of "us shaping our tools and our tools shaping us" happens in significantly shorter cycles as the internet rapidly evolves. Given the wide-ranging potential for middleware as a foundation for enabling user agency, it is important to distinguish between its near-term and long-term market development trajectories.

Markets for middleware-based services might be evaluated in terms of functional categories: quality curators, trust/reputation tracking services, harmfulness shields, and awfulness shields. They might alternately be considered in terms of who creates them: new specialist providers, or existing real-world organizations that decide to offer or partner in middleware services. There is significant potential value in journalism/news publishers, special interest communities, local communities and civic groups, academia and libraries, political groups, unions, and faith/values communities participating, as technical barriers to entry are reduced.

At least in the short term, it is important to distinguish between potential markets relating to two complementary visions for the future of middleware: community-focused and globally focused. Community-focused services favor a user base of cohesive, distinct, often small communities. This is a vision more in line with Mastodon's model, under which individual users set up and exert significant control their own instances, which largely reflect the preferences of distinct communities (or even just the individual themselves). At the same time, there are many globally focused services that are more open, public, and generally operate under a more permissionless model. Threads, for example, has integrated hundreds of millions of users into the broader ActivityPub "fediverse" but has a far less segmented or "opt-in" user experience than Mastodon, where users often select specific servers to join.[110] In the near term, these two visions and the various projects they represent are largely competing with one another for users and their attention.[111] But, as the market for middleware develops, there is promise that the various platforms, protocols, applications, and middleware services will begin to horizontally interoperate and complement one another more fully.[112]

---

[110] There is currently a debate within the Mastodon ecosystem regarding whether federation should be opt-out or opt-in, indicating that even within more community focused projects there is discussion around the extent to which separate communities should be linked.

Interoperability in the context of social media and middleware refers to the ability of different systems, platforms, and applications to seamlessly communicate, exchange, and use information. Interoperability can be horizontal, where different services or platforms work together seamlessly to allow cross-platform interaction, or it can be vertical where a platform gives third-parties access that enables them to build atop the platform. Interoperability allows users to transfer data, interact across various platforms, and utilize diverse services without being confined to a single ecosystem. In social media, horizontal interoperability ensures that users can connect and share content across different networks; middleware might facilitate this interaction by providing the necessary protocols and standards to bridge disparate systems. This has already become popular in the form of basic tools for multi-homing and cross-posting that provide a single interface that lets users interact across multiple platforms. As middleware services develop, increased horizontal interoperability should reduce the network effect advantage of large platforms, since even small services can enjoy higher network connectivity.

The long-term future of the social internet, therefore, might include a "pluriverse" where a few large platforms benefit from economies of scale while co-existing with numerous smaller community-focused services.

The future of middleware will remain very unevenly distributed in the near term. Initial market demand has been strongest from users wanting more granular control than they currently have; customers who paid for Block Party, for example, sought special protection from online harassment. People with a strong intolerance for spam or desire for custom curation mechanisms have historically been more willing to pay for special add-ons to help manage their inbox. However, the majority of internet users rarely change the default settings on their social media experiences; the value proposition of the middleware provider has to be worthwhile to overcome whatever friction cannot be reduced by good user experience design.

In the long run, we anticipate that globally focused platforms will see the value of opening themselves up to middleware to better manage content with satisfactory nuance, diversity, and subsidiarity. Policies promoting interoperability—particularly those that establish rights for users to delegate control to trusted agents or fiduciaries—could create opportunities for effective middleware services to thrive.[113] As those services demonstrate their effectiveness, they can increase awareness among markets and regulators, thus acting as a catalyst for broader middleware adoption and market success.[114]

---

[111] Helen Lewis, "The Weird, Fragmented World of Social Media After Twitter," *The Atlantic*, July 30, 2023, https://www.theatlantic.com/ideas/archive/2023/07/twitter-alternatives-bluesky-mastodon-threads/674859/.

[112] Richard Reisman and Chris Riley, "Community and Content Moderation in the Digital Public Hypersquare," *Tech Policy Press*, June 17, 2022, https://www.techpolicy.press/community-and-content-moderation-in-the-digital-public-hypersquare/.

# **Public Policy and Legal Frameworks**

The forces of the free market are necessary to bring the vision of middleware into being, but there is reason to doubt that they will be sufficient. Where markets are insufficient, public policy and regulation will be important components of a holistic approach to fostering a middleware ecosystem. After all, it has been official U.S. policy since 1996 that we should use law and public policy "to preserve the vibrant and competitive free market" for online services and "to encourage the development of technologies which maximize user control over what information is received by individuals."[115] In this section, we consider various levers of law and public policy that could be utilized to advance a market for middleware.

## Existing Regulatory Landscape Impacting Middleware

Since middleware providers undoubtedly handle user-generated content and might handle personal data, it is important for them to understand the ways existing laws and regulations apply to them. Here we examine the most pertinent laws and regulations for middleware services, starting with third-party liability, before discussing data privacy rules and competition regulations.

As noted above, perhaps no single piece of public policy has had a larger impact on the development of social media, and the internet more generally, than the law commonly referred to as Section 230.[116] Following a split of different courts over liability for content posted by users on two early online services, policymakers grew concerned about the potential chilling effect on free speech and the need for clear legal protection for internet service providers who wished to host and moderate user-created content without assuming liability. In order to address such concerns, Congress established Section 230 in 1996 as a legal shield for online platforms. It guarantees that sites which host user-generated content are not liable for the content of third parties except in certain narrow circumstances. As a result, platforms are able to perform basic editorial duties—such as removing obscene content, hate speech, and other lawful-but-awful

---

[113] Richard Whitt, *Reweaving the Web* (2024), https://www.reweavingtheweb.net/, "Hacking the SEAMs: Elevating Digital Autonomy and Agency for Humans," GLIA Foundation, August 8, 2020, https://papers.ssrn.com/sol3/papers.cfm?abstract_id=3669914.
[114] See John Hagel and Mark Singer, *Net Worth: Shaping Markets When Customers Make the Rules* (Cambridge: Harvard Business School Press, 1999).
[115] 47 U.S. Code § 230(b).
[116] 47 U.S. Code § 230 (1996).

content—without the risk that taking on such "editorial" roles would lead to liability for users' unlawful posts.[117]

The impact of Section 230 on the development of digital platforms has been profound. It has been credited with enabling the rapid growth and innovation of the internet economy by allowing platforms to scale without the threat of being directly liable for every piece of third-party content. The "Good Samaritan" provision, which protects platforms that take voluntary actions to restrict access to certain types of objectionable content, is an important reason that social media sites have been moderately effective in resisting a devolution into cesspools of horrible content.

However, as social media platforms have grown in influence and reach, Section 230 has come under increasing scrutiny, with critics arguing that protection from liability allows platforms to abdicate responsibility for censorship, misinformation, illegal activities, and harmful content disseminated on their services. For the past several years there have been steady calls to reform or repeal Section 230.[118] As society grapples with the balance between free speech, regulation, and accountability in the digital age, these debates will continue.

In the meantime, the question of how Section 230 applies is just as important to middleware providers as it is for incumbent platforms. Section 230 expressly protects not just platforms, but also the providers of technical tools that allow users to restrict access to unwanted material. This provision, designed to empower parents and other internet users to take more control over the content they saw online, anticipates the role that middleware providers may play. But there are open questions about the scope of this immunity, including whether it could protect middleware providers from being sued by platforms.

A novel legal theory was recently tested in the courts, exploring whether the law provides such protection from platform efforts to prevent use of middleware. A lawsuit filed on behalf of Ethan Zuckerman of the University of Massachusetts sought a federal court ruling to affirm that the law "protects the development of tools that empower social media users to control what they see online" from platforms' claims.[119] Professor Zuckerman intended to build his own software to enable users to automatically unfollow everyone on Facebook.[120] As the software was merely theoretical, the case was dismissed in November 2024, leaving the question unresolved. However, District Court Judge Jacqueline Scott Corley said that Prof. Zuckerman could refile the case in the future after he built the tool.[121] Had Zuckerman's argument prevailed, it might have established that middleware services could use Section 230 as a shield against lawsuits from platforms under various anti-hacking laws, such as the Computer Fraud and Abuse Act (discussed below). Although the case did not move forward, Zuckerman's argument highlights an important issue: policymakers should consider creating a statutory clarification that Section 230 protects third-party middleware providers from lawsuits by platforms.

---

[117] Daphne Keller and Oumou Ly, "The Breakdown: Daphne Keller explains the Communications Decency Act," Berkman-Klein Center for Internet and Society, August 13, 2020, https://www.youtube.com/watch?v=CkfXEtcSGyY.
[118] Valerie Brannon and Eric Holmes, Congressional Research Service, R46751, *Section 230: An Overview* (2024), https://crsreports.congress.gov/product/pdf/R/R46751.
[119] "Zuckerman v. Meta Platforms, Inc.," Knight First Amendment Institute at Columbia University, May 1, 2024, https://knightcolumbia.org/cases/zuckerman-v-meta-platforms-inc.
[120] David McCabe, "How a Law That Shields Big Tech Is Now Being Used Against It," *New York Times,* August 20, 2024, https://www.nytimes.com/2024/08/20/technology/meta-section-230-lawsuit.html.

Another important area of law and public policy that will have a direct impact on the development of middleware services is data privacy. Whether it be Europe's General Data Protection Regulation[122] (GDPR), various state laws such as the the California Consumer Privacy Act[123] (CCPA) and Utah Consumer Privacy Act[124] (UCPA), or federal proposals such as the recent American Privacy Rights Act[125] (APRA), data privacy laws and regulations govern how platforms and third parties collect, store, transfer, and (in some instances) process user data. Since middleware providers almost certainly handle user data, at least incidentally, these rules and regulations can both empower and restrict middleware services.

It is not within the scope of this paper to examine and dissect the myriad differences between all the laws that touch data privacy. Suffice to say that for European citizens, the GDPR imposes strict rules on data consent, access, and the right to be forgotten. The CCPA, itself largely modeled on the GDPR, provides similar protections to citizens of California. Taken together, the vast majority of commercially viable middleware will need to comply with the GDPR, the CCPA, or both, requiring middleware developers to design their products with privacy as a foundational element, potentially limiting how they collect and use data. This can affect middleware designed for analytics, advertising, and user experience enhancements, which typically rely on extensive data collection and processing.

On the federal level in the United States, proposed legislation such as the recent draft of the American Privacy Rights Act[126] (APRA) and the older American Data Privacy and Protection

---

[121] David McCabe, "Suit Against Meta, Using a Tech Shield Law, Is Dismissed," *New York Times,* November 7, 2024, https://www.nytimes.com/2024/11/07/technology/meta-section-230-lawsuit.html.
[122] Consolidated text: Regulation (EU) 2016/679 of the European Parliament and of the Council of 27 April 2016 on the protection of natural persons with regard to the processing of personal data and on the free movement of such data, and repealing Directive 95/46/EC (General Data Protection Regulation) (Text with EEA relevance), https://eur-lex.europa.eu/legal-content/EN/TXT/?uri=CELEX%3A02016R0679-20160504.
[123] California Consumer Privacy Act of 2018, (April 2024), https://cppa.ca.gov/regulations/pdf/cppa_act.pdf.
[124] Consumer Privacy Act, S.B. 227, 2022 General Session (2022), https://le.utah.gov/~2022/bills/static/SB0227.html.
[125] U.S. Senate Committee on Commerce, Science, and Transportation, "Committee Chairs Cantwell, McMorris Rodgers Unveil Historic Draft Comprehensive Data Privacy Legislation," April 7, 2024, https://www.commerce.senate.gov/2024/4/committee-chairs-cantwell-mcmorris-rodgers-unveil-historic-draft-comprehensive-data-privacy-legislation.
[126] Ibid.

Act[127] (ADPPA) aim to standardize privacy protections across states, which could simplify compliance for middleware developers but also impose new limitations on data practices. Such federal regulation would likely increase the importance of privacy-by-design principles and could necessitate significant adjustments in how middleware interacts with both data and platforms. Middleware solutions that enable enhanced privacy features or help platforms comply with these regulations could see increased demand.

Moreover, the development and enforcement of these privacy laws impact middleware by potentially limiting the type and volume of data that can be used, thus influencing the business models and operational strategies of middleware providers. These regulations make transparency, accountability, and user consent paramount, shifting how middleware platforms operate within the digital ecosystem. As a result, middleware that facilitates compliance and enhances user privacy could become a critical component of the digital infrastructure, aligning technical capabilities with regulatory requirements and public expectations for privacy and data protection. In order for this shift to take place, policymakers considering new data privacy rules should take the contextual nature of privacy into account and avoid crafting overbroad restrictions, as such rules could prevent middleware from accessing content or metadata in ways that might be beneficial.[128]

Finally, with regard to digital markets and the future development of third-party middleware services, the European Union's Digital Markets Act[129] (DMA) and Digital Services Act[130] (DSA) are perhaps the most significant regulatory frameworks established within the last five years. The DMA is primarily market focused, attempting to promote fairness and contestability in digital markets, while the DSA is primarily focused on societal impacts and safeguarding users' rights to freedom of information. Taken in conjunction with the GDPR, these regulations are a case study in the "Brussels Effect" and affect the development of technologies far beyond the continental shelf.[131]

---

[127] American Data Privacy and Protection Act, H.R. 8152, 117th Cong. § 2 (2022), https://www.congress.gov/bill/117th-congress/house-bill/8152/text.

[128] Richard Reisman, "How Third-Party Social Media Middleware Can Protect Contextual Privacy," *Tech Policy Press*, October 11, 2023, https://www.techpolicy.press/how-third-party-social-media-middleware-can-protect-contextual-privacy/.

[129] Regulation (EU) 2022/1925 of the European Parliament and of the Council of 14 September 2022 on contestable and fair markets in the digital sector and amending Directives (EU) 2019/1937 and (EU) 2020/1828 (Digital Markets Act) (Text with EEA relevance), https://eur-lex.europa.eu/eli/reg/2022/1925/oj.

[130] Regulation (EU) 2022/2065 of the European Parliament and of the Council of 19 October 2022 on a Single Market For Digital Services and amending Directive 2000/31/EC (Digital Services Act) (Text with EEA relevance), https://eur-lex.europa.eu/legal-content/EN/TXT/?uri=celex%3A32022R2065.

[131] The Brussels Effect generally refers to the European Union's ability to unilaterally shape global regulations and standards by leveraging its large consumer market, compelling multinational companies to adopt EU rules in their operations worldwide. See Anu Bradford, "The Brussels Effect," *Northwestern University Law Review,* vol. 107, no. 1 (2012), https://papers.ssrn.com/sol3/papers.cfm?abstract_id=2770634.

The DMA and DSA contain provisions that will significantly shape the future of third-party middleware, especially in how it interacts with major platforms. The DMA mandates that large online platforms ("gatekeepers") must ensure a higher degree of interoperability and portability with other platforms and third-party services. It also mandates that they provide access to their own services under fair, reasonable, and non-discriminatory conditions. A key provision requires gatekeepers to ensure that "end users or third parties authorised by end users can freely port the data continuously and in real time."[132] This could potentially lower barriers for middleware providers, enabling them to offer competitive services that leverage platform data and functionalities on behalf of the users they serve as agents. Importantly, however, within the context of data transfers, the DMA gives platforms some discretion to reject third parties that are deemed untrustworthy, which has spawned efforts to develop new methods for establishing trust between parties.[133] The DSA complements these measures by enforcing greater transparency in algorithmic processes and granting users more control over the content they see. This regulatory focus could encourage the development of middleware solutions that filter or curate content according to user preferences.

In the U.S., several proposals have been introduced that would, if passed, mandate more openness among social media platforms that might facilitate middleware services. One such proposal, dubbed the Augmenting Compatibility and Competition by Enabling Service Switching (ACCESS) Act, would require large platforms to allow users to delegate management of their account settings, content moderation, and online interactions to third-party agents.[134] Another proposal, New York State Senate Bill S6686, would go further still by specifying a right of delegation to middleware agent services in the form of open, standards-based API read and write access to a wide variety of data, and with explicit applicability to feeds/timelines and recommendations.[135] While these initiatives seem to currently be dormant, they are suggestive of how some in the United States are thinking about using public policy to support user agency in the form of rights of delegation, and could gain support as middleware gains recognition and begins to prove successful.

## First Amendment Issues Which Might Impact Middleware

Middleware can give users more autonomy as both listeners and speakers, which broadly serves the goals of the First Amendment. Platforms, however, might object that laws mandating middleware would infringe on the companies' *own* First Amendment rights. Platforms raised similar arguments in legal challenges to social media laws in Texas and Florida. In 2024, the Supreme Court in *Moody v. NetChoice* declined to fully resolve those cases, noting that the laws' so-called "must carry" mandates might potentially be justifiable in some situations, for example as applied to providers of email and direct messaging services. Nonetheless, the majority of justices indicated that for core social media services like the Facebook newsfeed, platforms indeed have the constitutional right to apply their own editorial policies without undue state interference.

---

[132] See Digital Markets Act.
[133] "Trust Model," Data Transfer Initiative, accessed December 3, 2024, https://dtinit.org/trust.
[134] ACCESS Act of 2019, S. 2658, 116th Cong. § 1 (2019), https://www.congress.gov/bill/116th-congress/senate-bill/2658.
[135] Senate Bill S6686, 2023–2024 Legislative Session (May 5, 2023), https://www.nysenate.gov/legislation/bills/2023/S6686.

How *Moody* might apply to a hypothetical law requiring interoperation with middleware is open for debate—and would of course depend on what that law actually said. Such a law would presumably, unlike the ones in *Moody*, leave platforms free to set and enforce their own editorial policies. They would simply have to offer their preferred moderation rules or ranking alongside other competing options for users. In that sense, such a law would interfere less with platforms' editorial rights than the Texas and Florida laws did. It would also avoid using state power to dictate specific new, content-based rules for online speech, as Texas and Florida did.

That said, a law making incumbent platforms interoperate with middleware providers might still unconstitutionally compel the platforms to "speak." For example, compelling platforms either to show users middleware service offerings, or to host unwanted content in order to make it available for curation by those competing providers, could be challenged on this basis. Some academics have argued that a law requiring only hosting (but not requiring platforms to show content in their own ranked feeds) could survive a First Amendment challenge.[136] Others have noted the Court's historical openness to laws that avoid imposing state-created rules for speech, but instead empower individual technology users to make their own decisions about content.[137]

## Deregulation Which Might Facilitate Middleware

There are several existing areas of law and public policy which inhibit the deployment of middleware. Most of these were well intentioned when first created but have failed to keep pace with the evolution of the internet; some have been abused to impede competition and innovation. Deregulation, meaning the repeal or reform of existing laws or regulations, will likely be a necessary part of any effort to promote a market for middleware development. Laws such as the Computer Fraud and Abuse Act, the Digital Millennium Copyright Act, and the Electronic Communications Privacy Act have aspects that hinder the viability of middleware. In this section, we discuss ways to address those aspects.

Under the Reagan administration, the rapid digitization of government—especially the Department of Defense and other national security agencies—led to fears that hostile actors would attempt to hack the government. After watching the 1986 movie *War Games,* in which a military supercomputer is inadvertently hacked by a high school student, President Reagan himself grew concerned and asked Congress to take action.[138] The product of all this was the Computer Fraud and Abuse Act (CFAA).

---

[136] Eugene Volokh, "Treating Social Media Platforms Like Common Carriers?," *Journal of Free Speech Law* vol. 377 (September 2021), https://papers.ssrn.com/sol3/papers.cfm?abstract_id=3913792.
[137] Francis Fukuyama, et al., *Amicus Curiae Brief in Support of Neither Party, Gonzalez v. Google LLC*, No. 22-277, December 7, 2023, https://www.supremecourt.gov/DocketPDF/22/22-277/292730/20231207155250803_22-277%20-555%20Fukuyama%20Amicus%20Brief%20Final.pdf.

The CFAA essentially makes it illegal to "knowingly [access] a computer without authorization" and establishes both criminal and civil penalties for doing so.[139] The criminal provisions of the CFAA have faced scrutiny from civil liberties activists for decades for various reasons, but it is the civil liabilities established by the CFAA that have been used to prevent the adversarial interoperation required for some middleware services when interoperation with those services is not actively enabled by the platforms, as is often the case. Most notably, Meta has repeatedly used the civil provisions of the CFAA to threaten middleware developers and successfully used the law to sue at least one company, Power Ventures, out of business for interoperating with Facebook.[140]

The Supreme Court recently narrowed the interpretation of what constitutes activity that exceeds authorized computer access, which is a positive development, but does not go far enough to allow the types of activity that may be necessary to develop a market for middleware.[141] Where platforms are unwilling to open their doors to third parties, it may be necessary for middleware providers to operate in an adversarial environment, and platforms should not be able to abuse the CFAA to prevent competition. Therefore, the civil provisions of the CFAA should be significantly narrowed to only allow platforms to seek damages when unauthorized access materially and significantly harms the platform, perhaps by raising the monetary threshold for civil claims. Even then, courts should dispose of these claims quickly—perhaps through some sort of tailored anti-SLAPP legislation—in order to prevent situations like the Power Ventures case, in which Facebook received *de minimis* compensation but was able to bleed Power Ventures dry through nearly a decade of litigation.

Another key law that should be considered is the Digital Millennium Copyright Act (DMCA). Signed into law by President Clinton in 1998, the DMCA is a massive piece of intellectual property legislation that was originally enacted in part in order to bring the U.S. into compliance with the World Intellectual Property Organization (WIPO) Treaties of 1996. It represents the federal government's attempt to reconcile intellectual property laws written for an analog age with the rise of digital technologies, and plays a pivotal role in shaping the landscape of online media. It is worth noting that Section 230 expressly carves out intellectual property law, so middleware providers must be aware that they must build products that comply with intellectual property laws and are responsive to rights holders. However, one section of the DMCA presents a serious challenge to the development of middleware: Section 1201.

---

[138] Fred Kaplan, "'WarGames' and Cybersecurity's Debt to a Hollywood Hack," *New York Times*, February 19, 2016, https://www.nytimes.com/2016/02/21/movies/wargames-and-cybersecuritys-debt-to-a-hollywood-hack.html.
[139] 18 U.S. Code § 1030(a)(1).
[140] *Facebook, Inc. v. Power Ventures, Inc.*, 828 F.3d 1068 (9th Cir. 2016), https://casetext.com/case/facebook-inc-v-power-ventures-inc.
[141] *Van Buren v. United States*, 141 S. Ct. 1648 (2021), https://www.supremecourt.gov/opinions/20pdf/19-783_k53l.pdf.

Promoted by the entertainment industry as a way to prevent piracy of copyrighted works, Section 1201 of the DMCA prohibits circumventing access controls—often referred to as digital rights management (DRM) technologies—that effectively protect copyrighted works.[142] It also makes it illegal to create, distribute, or market software or devices that are designed or produced primarily for the purpose of circumventing DRM technologies. In other words, Section 1201 makes it in many cases a federal crime to bypass, or help others bypass, any software intended to prevent access to a copyrighted work. Copyright holders, including platforms opposed to interoperating, can also bring civil claims.[143]

Middleware services that are designed to modify, filter, or enhance user access to digital content, and in doing so bypass any digital locks or encryption that restrict access to copyrighted materials on platforms, could potentially be seen as circumventing access controls. What's more, marketing or distributing such services for use by others could be a separate violation carrying with it jail time and large monetary penalties.

Some flexibility is already built into this law, since every three years the Copyright Office undergoes a review of exemptions to Section 1201. If granted, these last for three years, after which the applicant must seek a new exemption. Given that the Copyright Office has already granted exemptions that allow users to hack their own smartphones, smart TVs, and cars, as well as granted exemptions to security researchers and educators, it should also extend a Section 1201 exemption to third-party middleware providers and users of third-party middleware services. However, seeking and receiving a Section 1201 exemption every three years is a laborious and costly process and middleware providers are not guaranteed. Therefore, while a triennial exemption from Section 1201 for middleware providers would be beneficial, a far better and more long-term solution would be for Congress to establish a statutory exemption for middleware.

Finally, we have the Electronic Communications Privacy Act (ECPA) of 1986, which combined several different legislative proposals into one vehicle in order to extend government restrictions on wiretaps from telephone calls to include transmissions of electronic data by computer.[144] The ECPA is divided into three parts: the Wiretap Act, which prohibits the interception of real-time communications; the Stored Communications Act, which deals with the privacy of communications held in electronic storage; and the Pen Register Act, which restricts the recording of dialing, routing, and signaling information.

---

[142] 17 U.S. Code § 1201.
[143] Shreyanka Mirchandani Changaroth and Lateef Mtima, "What is Section 1201 Digital Millennium Copyright Act?: A Legislative Primer," Institute for Intellectual Property and Social Justice, https://iipsj.org/wp-content/uploads/2023/09/Section-1201-Legislative-Primer.pdf.
[144] 18 U.S. Code Chapter 119.

Middleware services, which act as intermediaries that modify or manage data between networks or applications, can potentially implicate various aspects of the ECPA, particularly if these services are used to intercept, store, or modify electronic communications without proper authorization. For instance, a middleware service that caches or routes communications might come under scrutiny if it inadvertently stores or reveals contents without user consent, potentially breaching provisions of the Stored Communications Act.[145] Furthermore, if middleware involves real-time processing or altering of communications, this could be interpreted as interception, which is tightly regulated under the Wiretap Act. Thus, middleware developers need to be acutely aware of the ECPA's requirements, ensuring that their services operate within the bounds of the law while handling electronic communications.

Considering the complexities introduced by middleware services and their potential to run afoul of ECPA, legislative updates could be beneficial. One approach could be to clarify the definitions and exceptions in the ECPA specifically for middleware technologies, ensuring that they can operate effectively without unintentional legal violations. Moreover, updating the ECPA to explicitly accommodate modern digital communication practices and technologies could foster innovation while maintaining robust privacy protections. This might involve expanding the lawful consent exceptions, clarifying what constitutes consent for information that is related to but not controlled or owned by the principal, and refining the terms under which middleware can lawfully interact with electronic communications, thus providing a safer legal framework for middleware developers to innovate and expand their services.

## New Authorities and Existing Agencies Which Might Facilitate Middleware

Even with significant statutory change, our regulatory agencies themselves may still be ill equipped to fully understand the socioeconomic and technical nuances of digital markets for middleware.The creation of a thriving market for middleware may require regulators to be granted new authorities or utilize existing authorities in creative ways. In particular, the Stanford Working Group on Platform Scale highlighted the need for new authorities to govern the "availability of platform APIs to middleware providers, platform compliance with other conditions necessary to allow middleware providers to offer their products, and fair revenue sharing and adherence to rules that allow middleware business models to thrive."[146] They also noted the more challenging need for regulators, industry leaders, and middleware developers to design and implement both regulatory and technical frameworks to facilitate the market for middleware.

---

[145] Orin Kerr, "A User's Guide to the Stored Communications Act, and a Legislator's Guide to Amending It," *72 George Washington Law Review,* vol. 72, no. 6 (last revised July 6, 2016), https://papers.ssrn.com/sol3/papers.cfm?abstract_id=421860.
[146] See Fukuyama et al., "Report of the Working Group on Platform Scale."

There are several approaches that policymakers could take to address these challenges. The most ambitious and comprehensive approach would be to establish a new, specialized regulatory agency with the primary responsibility and goal of fostering a middleware market, along with other issues related to platform power and competition. A slightly less ambitious approach would be to empower an existing agency or group of agencies with new authorities. Finally, there are existing authorities and soft-law mechanisms which could be leveraged to foster a middleware market. We will address the benefits and drawbacks of each of these approaches in turn.

As it stands, jurisdiction over the public policy related to middleware spans numerous federal agencies, including the Federal Trade Commission (FTC), Federal Communications Commission (FCC), Department of Commerce (DOC), Consumer Financial Protection Bureau (CFPB), U.S. Patent and Trademark Office (USPTO), and the U.S. Copyright Office (USCO). While the Office of Science and Technology Policy (OSTP) at the White House aims to develop and implement unified strategies for federal policies related to technology, the reality is that these numerous agencies have differing, and oftentimes conflicting, objectives, cultures, and mandates. This fragmentation can lead to inconsistent policy application, creating a regulatory environment that is complex and challenging for middleware companies to navigate. The lack of a centralized authority or coherent strategy specifically tailored for middleware not only hampers the efficiency of regulatory processes but also impedes the sector's ability to innovate and grow. With middleware playing an increasingly important role in enhancing interoperability and competition within the digital ecosystem, a more unified regulatory approach could streamline compliance, reduce uncertainty, and foster an environment that better supports the development of middleware solutions as well as the ecosystem of platforms and other systems they connect.

Creating a new regulatory agency specifically designed to foster a market for middleware is the most comprehensive, though perhaps least feasible, way to address this challenge. A singular agency with expertise around and authority over the market for middleware has the primary benefit of having a unity of vision.[147] Bringing together technologists, economists, lawyers, and policy experts under one roof with a singular purpose and vision would unquestionably enhance the ability to tailor regulations that specifically address the unique needs and challenges of middleware technologies. This could include setting standards for interoperability and data portability between dominant digital platforms and middleware providers, ensuring that middleware solutions are developed and deployed in ways that respect user privacy and data security, and maintaining a balance between innovation and consumer protection.

---

[147] Harold Feld, "The Case for the Digital Platform Act: Market Structure and Regulation of Digital Platforms," Public Knowledge, May 2019, https://publicknowledge.org/policy/the-case-for-the-digital-platform-act/.

However, the establishment of a new regulatory body also presents unique challenges and drawbacks. As Ronald Reagan warned, "The first rule of a bureaucracy is to protect the bureaucracy." Federal agencies are notoriously territorial, and it is unlikely that the numerous agencies whose jurisdiction would be affected or curtailed by a new agency would voluntarily give up their authority. There's also the risk of regulatory overlap and confusion if the roles and responsibilities of the new agency aren't clearly delineated from those of existing bodies such as the FTC or FCC. Furthermore, the process of setting up such an agency would likely be complex and resource-intensive, requiring significant time, financial investment, and political capital that might otherwise go to other political priorities.

In any case, the effectiveness of a new regulatory agency would heavily depend on its ability to adapt and respond to the rapidly evolving tech landscape. Middleware technologies and the digital market as a whole are characterized by rapid innovation and change. A regulatory framework that is too rigid or slow to adapt could quickly become obsolete, failing to address new technological developments or changing market conditions. Therefore, while the idea of a specialized agency to foster a middleware market is promising, it requires careful planning and flexibility in its regulatory approach to truly benefit the ecosystem without hampering growth or innovation.

The second option—granting existing agencies new authorities, or clarifying how they should utilize existing authorities to foster a market for middleware—may be more feasible. This approach leverages existing resources and expertise while potentially filling regulatory gaps specifically related to middleware.[148] For instance, the FCC, which already oversees broadly analogous interoperability mandates in the telecommunications space, could bring relevant experience to the task. Or the FTC, which already oversees consumer protection and antitrust issues, could be granted enhanced powers to address anti-competitive practices that affect middleware companies, particularly in enforcing open access to platform APIs. The Department of Commerce could support middleware development through innovation-promoting policies and international trade negotiations that protect U.S. middleware companies abroad.

In choosing to lean on existing agencies to ensure that a market for middleware is free and fair, policymakers should focus on the FTC and FCC, which already have broad authority over many areas related to technology and social media. In fact, absent specific congressional directives, a friendly FTC could unilaterally establish new rules around data privacy, data security, platform openness, algorithmic transparency, and numerous other issues that are related to middleware development. Indeed, the FTC is unilaterally working to establish new rules around data privacy and security.[149] For their part, some at the FCC are already looking at ways the Commission can leverage existing authorities to address actions by platforms to prevent third-party services, particularly with regards to messaging services.[150]

---

[148] "Stigler Committee on Digital Platforms: Final Report," Stigler Center for the Study of the Economy and the State, University of Chicago Booth School of Business, September 16, 2019, https://www.chicagobooth.edu/research/stigler/news-and-media/committee-on-digital-platforms-final-report/.

Two major obstacles to the FTC taking actions that might support a market for middleware is the strenuousness of the Magnuson-Moss rulemaking process[151] and inevitable court challenges over new regulations, including objections under the First Amendment. Clear congressional directives are always preferable to unilateral regulatory actions. Another inhibition is the lack of technical expertise at the FTC. Primarily composed of lawyers and economists, the FTC recently established a new Office of Technology with the purpose of helping the Commission "keep pace with technological challenges in the digital marketplace," by having technologists advise on issues related to digital markets.[152] Should policymakers decide to give the FTC new authorities and responsibilities related to middleware, they will also need to ensure that the Commission has the resources and expertise needed to handle such responsibility.

Antitrust enforcement authorities at the FTC and Department of Justice Antitrust Division represent another avenue for fostering a thriving middleware ecosystem. By actively scrutinizing the competitive practices of dominant platforms, these agencies can dismantle barriers that inhibit middleware providers from entering or expanding in the market. Targeted enforcement actions could address anticompetitive behaviors such as exclusive dealing or self-preferencing that prevent middleware from gaining traction. Moreover, antitrust remedies—like requiring interoperability or curbing platform control over proprietary algorithms—could create fertile ground for middleware innovation. By emphasizing competitive fairness, these enforcement efforts align with the broader goals of decentralization and user empowerment, ensuring that middleware can thrive as a viable alternative to centralized platform dominance.

The major obstacle facing the FCC regarding actions that might support a market for middleware is the lack of clear regulatory authority over platforms. As it stands, online platforms are generally categorized as "information services" under the Telecommunications Act of 1996 and are thus exempt from many of the FCC's strongest regulatory powers. Of course, the distinct nature of digital platforms means that simply reclassifying platforms as Title II telecommunications services and treating them the same as common carriers poses constitutional and practical challenges.¹⁵³ Ideally, Congress would revisit the entire Telecommunications Act and perhaps establish new titles to give the FCC more specific authority over digital platforms. In the meantime, the FCC could look at existing authorities such as the Commissions' Part 14 rules to determine the extent to which platforms' blocking of third-party services may impede accessibility and useability.

---

[149] "FTC Releases 2023 Privacy and Data Security Update," Federal Trade Commission, March 28, 2024, https://www.ftc.gov/news-events/news/press-releases/2024/03/ftc-releases-2023-privacy-data-security-update.
[150] https://www.theverge.com/2024/2/12/24071226/fcc-commissioner-brendan-carr-apple-beeper-mini
[151] Dan Bosch, "Primer: The FTC and Magnuson-Moss Rulemaking," American Action Forum, September 21, 2022, https://www.americanactionforum.org/insight/primer-the-ftc-and-magnuson-moss-rulemaking/.
[152] "FTC Launches New Office of Technology to Bolster Agency's Work," Federal Trade Commission, February 17, 2023, https://www.ftc.gov/news-events/news/press-releases/2023/02/ftc-launches-new-office-technology-bolster-agencys-work.

Expanding the authority of existing agencies also comes with challenges. Each agency has its established priorities and operational cultures, which may not seamlessly integrate the specific needs of middleware development. There is a risk of diluting focus or of having insufficient specialization in the complex technical and economic issues unique to middleware. Moreover, inter-agency coordination can be problematic, given overlapping jurisdictions and the potential for bureaucratic inertia. For middleware, which requires agile and informed regulatory responses to keep pace with rapid technological advancements, these bureaucratic delays could stifle innovation rather than foster it.

Therefore, while empowering existing agencies avoids the complexities of establishing a new entity, it requires careful consideration of how these expanded powers are implemented and coordinated. Policymakers would need to ensure that these agencies are equipped with not only the necessary legal authority, but also the resources and specific mandate to address middleware issues. This approach demands a concerted effort to enhance inter-agency collaboration, streamline regulatory processes, and, perhaps most importantly, maintain an ongoing dialogue with both the middleware industry and the broader tech community to stay aligned with the sector's evolution and needs.

## Soft Law and Voluntary Standards

Given industry's lackadaisical compliance with European regulations around openness, interoperability, and data portability, it must be noted that a purely adversarial regulatory approach to fostering a market for middleware may not be the most productive avenue. While there are certainly areas of public policy, such as data privacy, that will require policymakers to create new rules, they should also consider a carrot-and-stick approach. What's more, having non-technical agencies establish hard technical standards can inadvertently stifle innovation and competition, as we have seen with the impact of the FTC's implementation of the Children's Online Privacy Protection Act on the development of privacy-affirming technologies.¹⁵⁴ Instead of crafting hard rules about every aspect of how regulators wish to see a market for middleware develop, it may prove more effective to utilize soft law.

---

¹⁵³ Daphne Keller, "The Long Reach of Taamneh: Carriage and Removal Requirements for Internet Platforms," Brookings Institution, October 19, 2023, https://www.brookings.edu/articles/the-long-reach-of-taamneh-carriage-and-removal-requirements-for-internet-platforms/.

¹⁵⁴ Neil Chilson, "Case-by-Case! Old Statutes and New Tech at the FTC," in *Rulemaking Authority of the US Federal Trade Commission*, Daniel A Crane, ed. (Paris: Concurrences, 2022).

More commonly applied to the field of international relations and lacking an accepted definition, soft law refers, for our purposes, to mechanisms which have no strict legal force but are influential in guiding behavior and shaping norms. Since its inception, the internet has relied heavily on soft-law mechanisms, primarily in the form of technical standards. The benefit of employing soft law in promoting middleware development lies in its adaptability and collaborative nature. Soft law allows for ongoing modifications and updates in response to technological advancements and market needs, which is vital in the fast-evolving tech landscape. By engaging industry stakeholders in the creation of these standards, policymakers can ensure that the guidelines are practical and implementable. This participatory approach can lead to greater buy-in from major platforms, which is crucial for the effective implementation of any regulatory strategy.

One recent example of how soft law can guide industry behavior while encouraging innovation and maintaining compliance with overarching principles is the National Institute for Standards and Technology (NIST)'s AI Risk Management Framework (AI RMF).[155] Designed as a voluntary guideline to assist organizations in managing risks associated with artificial intelligence systems, this framework is comprehensive and flexible, suitable for various sectors aiming to harness AI technologies responsibly. While compliance with the AI RMF is voluntary, its principles align closely with emerging regulatory expectations around AI globally, such as those related to fairness, transparency, and privacy. As a result, firms are incentivized to comply with the AI RMF, as it can minimize the need for major overhauls of AI systems to meet new legal standards. Furthermore, demonstrating compliance with an established and respected framework like the NIST AI RMF can serve as evidence of good-faith efforts to adhere to best practices, potentially mitigating regulatory scrutiny and penalties. The result is that most firms are voluntarily complying with the AI RMF, and thus reducing risks associated with AI, not because of hard laws but because of an alignment of incentives. Such a soft-law approach to managing AI risks has the added benefit of being far more flexible than a hard regulation or mandatory standard, since firms can be more creative in how they choose to comply with the framework and NIST has the ability to continually tweak the framework as new innovations occur.

While the NIST AI RMF provides a useful example of soft law in guiding responsible technology practices, NIST may not be the best venue for developing the specific interoperability standards required for middleware. Standards development organizations (SDOs) like the Internet Engineering Task Force (IETF) and the World Wide Web Consortium (W3C) have decades of experience crafting technology-specific standards that underpin the internet, including protocols and frameworks central to social networking and interoperability, and are perhaps better suited to standards development in these technologies. Their established processes and deep technical expertise make them well equipped to develop and implement the kinds of interoperability standards needed to support middleware, potentially providing a more effective venue than NIST for this critical work. The primary downsides of using SDOs as opposed to a government agency such as NIST is that SDOs often face slow development cycles due to their consensus-driven processes, delaying the implementation of critical interoperability frameworks for middleware. Additionally, their reliance on voluntary adoption and the disproportionate influence of large industry players can lead to limited enforcement mechanisms and standards that favor incumbents over fostering competition and innovation.

---

[155] NIST, Department of Commerce, NIST AI 100-1, *Artificial Intelligence Risk Management Framework (AI RMF 1.0),* (January 2023), https://doi.org/10.6028/NIST.AI.100-1.

When it comes to establishing regulatory frameworks that support a market for middleware, there are numerous areas where a soft-law approach might be preferable to new regulations or authorities. One area where soft-law standards could be incorporated into a unified agenda for governing and regulating social media is API openness. As we have noted, a purely adversarial approach to middleware development is unlikely to succeed, and there must be platform buy-in. Embedding hard, technical rules about exactly how platforms must allow third-party access creates clarity, but is also inflexible, potentially inhibiting innovation and even hampering compliance with the spirit, if not the letter, of the law.

A better approach to encourage platforms to open their APIs could be to have NIST establish a framework for platform openness, outlining both broad principles and technical standards to promote and enable more interoperability and better data portability. If policymakers wish to further incentivize platforms to comply with such a framework, they could choose to tie compliance with the framework to other aspects of law and public policy. Through either hard statute or softer memorandums, for example, compliance with an openness framework could be a consideration for FTC enforcement actions, or it could be incentivized through provisions in a federal data privacy law.

## The Need for Safe Harbors

It is important to note that incumbent platforms are likely to resist any public policy changes that require or otherwise encourage them to open up their ecosystems. One reason platforms have cited in opposing past efforts at inducing openness is that, without legal safe harbors, they may find themselves vulnerable to lawsuits for actions taken by third-party middleware services operating within their systems. It is important to consider whether and what sort of legal protections incumbent platforms may be needed in order to facilitate a more open and competitive digital ecosystem.

A historical precedent illustrating this concern is the aftermath of the Cambridge Analytica scandal that engulfed Facebook. In 2018, it was revealed that British political consulting firm Cambridge Analytica had harvested personal data from millions of Facebook profiles without users' consent and in contravention of Facebook's policies, and used that data for targeted political advertising. This scandal resulted in significant public backlash and legal scrutiny. Facebook faced numerous lawsuits, and a $5 billion fine from the FTC. While the Cambridge Analytica scandal is complex, and Facebook is not blameless, the incident highlights the legal risks that platforms face when third parties misuse their data. Without legal assurances that limit liability, platforms will remain reluctant to integrate third-party services, fearing similar repercussions.

Incumbent platforms regularly argue that opening themselves up to third-party middleware services could dilute their control over user data and engagement, making it more difficult for them to comply with data privacy rules, intellectual property laws, and other legal requirements. Extending protection to cover interoperation with third-party middleware services would encourage platforms to open their ecosystems without fear of being held liable for every action taken by these services.

# **Conclusion**

Just as open markets for personal computing software and app stores for smartphones enabled once-speculative markets to explode, middleware holds transformative potential to foundationally reshape the social media landscape. By decentralizing control and introducing a competitive layer of third-party services, middleware empowers users to take charge of their online experiences, reducing the dominance of centralized platforms and the power they hold over public discourse. This shift has the potential to not only enhance user agency but also spark innovation, paving the way for more inclusive and engaging online spaces. At the same time, it also raises complex questions about polarization, user safety, and cross-platform privacy.

Middleware addresses several critical challenges with current social media platforms, including legitimacy issues around content moderation and concern about opaque private control of content curation—and, by extension, the integrity of public discourse. However, realizing the full promise of middleware requires thoughtful design, strategic market alignment, and supportive regulatory frameworks. The development of viable business models, whether for-profit, nonprofit, or hybrid, is essential for the sustainability of middleware services. Additionally, policies that promote platform openness, interoperability, and user rights will be crucial in fostering a thriving middleware ecosystem. Addressing regulatory barriers and encouraging innovation through targeted deregulation and potentially through new authorities can further support the growth and integration of middleware to provide an expanding variety of services in new and unanticipated ways.

Middleware offers a compelling solution to pressing issues of platform power and user agency in the digital realm. By empowering users and fostering a competitive market for content curation and moderation services of all kinds, it has the potential to create a more balanced, inclusive, and democratic online environment. That path forward will require a whole-of-society collaboration among technologists, policymakers, platforms, and users to ensure that middleware fulfills its potential to reshape social media—and thus public discourse—for the better.